\documentclass[sigconf,balance=false]{acmart}
\usepackage{popets}
\setcopyright{popets}
\copyrightyear{YYYY}

\acmYear{YYYY}
\acmVolume{YYYY}
\acmNumber{X}
\acmDOI{XXXXXXX.XXXXXXX}
\acmISBN{}
\acmConference{Proceedings on Privacy Enhancing Technologies}
\usepackage{tikz}
\usepackage{amsmath}

\usepackage{enumitem}
\usepackage{bbding}
\usepackage{float}
\usepackage{caption}
\usepackage{subcaption}
\usepackage{censor}
\usepackage{url}
\usepackage{multirow}
\usepackage{graphicx}
\usepackage{graphicx}
\usepackage{comment}

\begin{document}

\title[Exploring the Privacy Risks of Adversarial VR Game Design]{Exploring the Privacy Risks of Adversarial VR Game Design}

\author{Vivek Nair$^*$}
\affiliation{%
  \institution{UC Berkeley}
  \city{Berkeley}
  \state{CA}
  \country{USA}
}
\email{vcn@berkeley.edu}

\author{Gonzalo M. Garrido$^*$}
\affiliation{%
  \institution{TU Munich}
  \city{Munich}
  \state{}
  \country{Germany}}
\email{gon.munillag@gmail.com}

\author{Dawn Song}
\affiliation{%
  \institution{UC Berkeley}
  \city{Berkeley}
  \state{CA}
  \country{USA}}
\email{dawnsong@berkeley.edu}

\author{James F. O'Brien}
\affiliation{%
  \institution{UC Berkeley}
  \city{Berkeley}
  \state{CA}
  \country{USA}}
\email{job@berkeley.edu}

\renewcommand{\shortauthors}{Nair et al.}

\begin{abstract}
Fifty study participants playtested an innocent-looking ``escape room'' game in virtual reality (VR). Within just a few minutes, an adversarial program had accurately inferred over 25 of their personal data attributes, from anthropometrics like height and wingspan to demographics like age and gender.
As notoriously data-hungry companies become increasingly involved in VR development, this experimental scenario may soon represent a typical VR user experience.
Since the Cambridge Analytica scandal of 2018, adversarially-designed gamified elements have been known to constitute a significant privacy threat in conventional social platforms. In this work, we present a case study of how metaverse environments can similarly be adversarially constructed to covertly infer dozens of personal data attributes from seemingly-anonymous users.
While existing VR privacy research largely focuses on passive observation, we argue that because individuals subconsciously reveal personal information via their motion in response to specific stimuli, active attacks pose an outsized risk in VR environments.

\end{abstract}

\keywords{virtual reality, metaverse, data harvesting, privacy, anonymity}

\maketitle

\def\thefootnote{*}\footnotetext{Equal contribution.}

\begin{table*}[!ht]
\resizebox{\textwidth}{!}{%
\begin{tabular}{l|lcccc|lcccc|}
\cline{2-11}
 & \multicolumn{5}{c|}{\textbf{Data Sources}} & \multicolumn{5}{c|}{\textbf{Observable Attribute Classes}} \\ \hline
\multicolumn{1}{|c|}{\textbf{Attacker Type}} & \multicolumn{1}{c|}{\textbf{\begin{tabular}[c]{@{}c@{}}Raw Sensor\\ Data\end{tabular}}} & \multicolumn{1}{c|}{\textbf{\begin{tabular}[c]{@{}c@{}}Processed\\ Telemetry\end{tabular}}} & \multicolumn{1}{c|}{\textbf{\begin{tabular}[c]{@{}c@{}}Rendering Pipeline\\ \& Host System APIs\end{tabular}}} & \multicolumn{1}{c|}{\textbf{\begin{tabular}[c]{@{}c@{}}Networked\\ Telemetry\end{tabular}}} & \textbf{\begin{tabular}[c]{@{}c@{}}Presented\\ Telemetry\end{tabular}} & \multicolumn{1}{l|}{\textbf{Device}} & \multicolumn{1}{l|}{\textbf{Network}} & \multicolumn{1}{l|}{\textbf{Geospatial}} & \multicolumn{1}{l|}{\textbf{Audio}} & \multicolumn{1}{l|}{\textbf{Behavior}} \\ \hline
\multicolumn{1}{|l|}{\textbf{Privileged Attacker I}} & \multicolumn{1}{c|}{\Checkmark} & \multicolumn{1}{c|}{\Checkmark} & \multicolumn{1}{c|}{\Checkmark} & \multicolumn{1}{c|}{} &  & \multicolumn{1}{c|}{\Checkmark} & \multicolumn{1}{c|}{} & \multicolumn{1}{c|}{\Checkmark} & \multicolumn{1}{c|}{\Checkmark} & \Checkmark \\ \hline
\multicolumn{1}{|l|}{\textbf{Privileged Attacker II}} & \multicolumn{1}{l|}{} & \multicolumn{1}{c|}{\Checkmark} & \multicolumn{1}{c|}{\Checkmark} & \multicolumn{1}{c|}{\Checkmark} &  & \multicolumn{1}{c|}{\Checkmark} & \multicolumn{1}{c|}{\Checkmark} & \multicolumn{1}{c|}{\Checkmark*} & \multicolumn{1}{c|}{\Checkmark} & \Checkmark \\ \hline
\multicolumn{1}{|l|}{\textbf{Privileged Attacker III}} & \multicolumn{1}{l|}{} & \multicolumn{1}{l|}{} & \multicolumn{1}{l|}{} & \multicolumn{1}{c|}{\Checkmark} & \Checkmark & \multicolumn{1}{l|}{} & \multicolumn{1}{c|}{\Checkmark} & \multicolumn{1}{c|}{\Checkmark*} & \multicolumn{1}{c|}{\Checkmark*} & \Checkmark \\ \hline
\multicolumn{1}{|l|}{\textbf{Non-Privileged Attacker}} & \multicolumn{1}{l|}{} & \multicolumn{1}{l|}{} & \multicolumn{1}{l|}{} & \multicolumn{1}{l|}{} & \Checkmark & \multicolumn{1}{l|}{} & \multicolumn{1}{c|}{} & \multicolumn{1}{c|}{\Checkmark*} & \multicolumn{1}{c|}{\Checkmark*} & \Checkmark \\ \hline
\end{tabular}%
}
\centering{\vspace{0em}\\ *Observable only in weaker filtered/preprocessed format}
\caption{Virtual reality threat actor capabilities.}
\label{tab:threat-model}
\vspace{-1em}
\end{table*}
\section{Introduction}
\label{sec:Introduction}

Through the fog of rapidly shifting preferences for internet platforms, one clear trend has stood the test of time: with each new and improved medium for accessing the web comes a new and improved method for harvesting personal user data. As these technologies become more immersive and tightly integrated with our daily lives, so too do the corresponding intrusive attacks on user privacy.

Privacy concerns have existed since the early days of the world wide web, even when users primarily accessed information through static websites. The emergence of social media platforms in the early 2000s changed this paradigm, generating a torrent of additional data on user behavior. Third-party (tracking) cookies that can uniquely identify and follow individuals~\cite{cookies} around the web then allowed this data to be deployed for everything from surveillance advertisement~\cite{surveillance_ads} to pushing political agendas~\cite{political_targeting}.

In the past decade, users shifted to accessing the web primarily via their mobile phones ($92.1\%$ as of 2022~\cite{www_users}), simultaneously introducing a suite of newly-extractable data attributes like audio, video, and geolocation. Next, the wave of wearable devices such as smart watches added critically sensitive data like biometrics and health information into the mix~\cite{facebook_fine}. Most recently, virtual home assistants have made possible pernicious intrusions into users' most private activities~\cite{edu_smart_2021}. Overall, the tendency is clear: each new technology has gradually expanded the scope of data attributes accessible to attackers, further eroding any expectation of privacy.

Virtual reality (VR) is well-positioned to become a natural continuation of this trend. While VR devices have been around in some form since well before the internet~\cite{vr_history}, the true ambition of major corporations to turn these devices into massively-connected social ``metaverse" platforms has only recently come to light~\cite{metaverse_intro, Blackrock_meta, Morgan_meta}.
These platforms, by their very nature, turn every gaze, movement, and utterance of a user into a stream of data, broadcast to other users around the world in the name of facilitating real-time interaction.

It has long been understood that individuals exhibit distinct biomechanical motion patterns that can be used to identify them or infer their personal attributes \cite{Obrien:2000:AJP, Kirk:2005:SPE, cutting_recognizing_1977, kozlowski_recognizing_1977, pollick_gender_2005, jain_is_2016}, which researchers have shown can be exploited to identify and profile users in VR \cite{https://doi.org/10.48550/arxiv.2302.08927, https://doi.org/10.48550/arxiv.2301.05940, miller_personal_2020, 10.1145/3411764.3445528}. While existing work has largely focused on passive observation of VR users, the success of games specifically designed to harvest user data \cite{8436400} on conventional social platforms motivates us to investigate similar active attacks in VR.

Gaming has been the predominant driver of the recent uptick in VR adoption, with 91 of the 100 most popular VR applications being games \cite{steam_most_used}.
This paper aims to shed light on the significant privacy risks associated with adversarially-designed VR games that appear innocuous to end users.
We have identified over $25$ examples of private data attributes that attackers can covertly harvest from VR users, which we experimentally demonstrate in our $50$-person user study.
Many of these attributes would be difficult to observe passively but can be obtained with high fidelity by prompting users to unknowingly reveal more information about themselves via carefully designed interactive game elements.
We aim to increase broad awareness of privacy concerns within VR and encourage privacy practitioners to examine the challenges and solutions that lie at the intersection of data privacy and the VR-enhanced social internet. \\

\noindent The main contributions of our study are:
\medskip
\begin{enumerate}[leftmargin=*]
    \itemsep 0.5em
    \item We provide a comprehensive framework of virtual reality information flow, threat actors, data sources, observable attribute classes (\S\ref{sec:threat_model}), and systematic privacy attacks (\S\ref{sec:Privacy_attacks}).
    \item With our open-source VR demo, we illustrate how malicious game developers could design adversarial yet seemingly-innocuous VR environments that trick users into revealing personal information through their motion and behavior patterns (\S\ref{sec:Experimental_Design}).
    \item We experimentally demonstrate how an attacker can covertly harvest over $25$ unique data attributes from VR users, many of which are difficult to obtain through passive observation (\S\ref{sec:Experiment_results}).
\end{enumerate}

\eject

\section{Background}
\label{sec:threat_model}



The aim of this section is to provide background information on virtual reality devices and applications necessary to understand the substance of this paper. We outline a holistic framework of VR information flow and threat actors by which the current contributions can be clearly differentiated from the existing literature.

\subsection{VR Information Flow}

A VR device uses an array of external or onboard sensors to generate a stream of information about its user. In a typical consumer-grade VR system, the position and orientation of a head-mounted display (HMD) and two hand-held controllers are measured in 3D space (six degrees of freedom per tracked object) at a rate of between 60~Hz and 144~Hz, resulting in a “telemetry stream."

\begin{figure}[!ht]
    \centering
    \includegraphics[width=\linewidth]{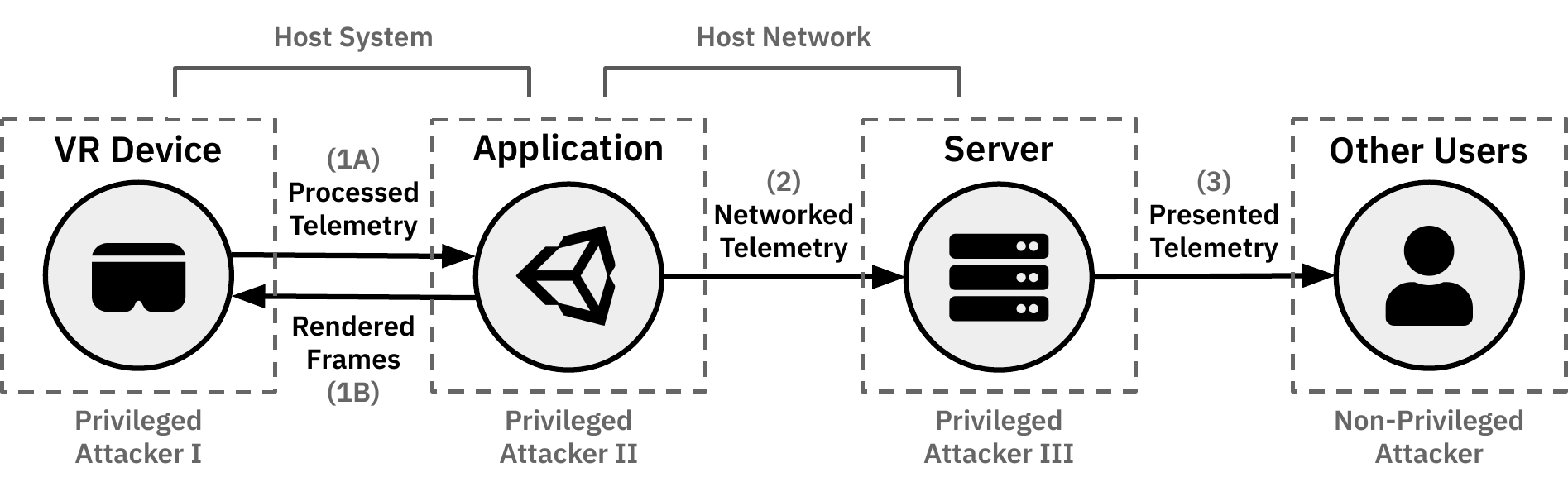}
    \vspace{-2em}
    \caption{Virtual reality information flow and threat model.}
    \label{fig:system}
\end{figure}

Users can download various games and applications for their VR device from an app store provided by the manufacturer. These applications consume the telemetry stream generated by the VR device and use it to render stimuli for the user, thereby creating an immersive experience. In the case of virtual telepresence (or “metaverse”) applications, the telemetry data is also shared with a variety of external systems. The typical information flow for such an application, as depicted in Fig.~\ref{fig:system}, is as follows:

\begin{enumerate}[leftmargin=*]
    \item The VR device processes raw sensor data into useful telemetry, which it provides to the application via an API. The application uses this data to provide stimuli to the user via a rendering pipeline, which the application completely controls.
    \item If the application involves interactions with other users, the client-side VR application streams processed telemetry data to an external server via a network to facilitate such interactions.
    \item The server then relays this data to other users so that an ``avatar'' of the original user can be rendered on their devices.
\end{enumerate}

\vspace{-0.5em}

\vspace{-0.5em}

\noindent The general basis of the VR privacy threat is that telemetry information available to an adversary can be used to infer private user data instead of being used for its intended purpose.
Accordingly, our VR threat model considers four potential adversaries corresponding to the four distinct entities associated with the VR information flow. We summarize the capabilities of each attacker in Table~\ref{tab:threat-model}.



\subsection{VR Threat Model}
\medskip

\noindent \textbf{Privileged Attacker I} (the ``Hardware Adversary'').\\
The first privileged attacker represents the party controlling the firmware of a target user's VR device. This attacker has access to raw sensor data from the VR device, including spatial telemetry, audio/visual streams, and device specifications. There is a bi-directional information flow between the device and the local application: the device provides processed telemetry to a running application, which the attacker can manipulate arbitrarily, and the device receives a stream of audio/visual stimuli from the application, which the attacker can manipulate arbitrarily before presenting to the user. However, this attacker cannot read or manipulate the network communications of the application.\medskip

\noindent \textbf{Privileged Attacker II} (the ``Client Adversary'').\\
Our second privileged attacker represents the developer of the client-side VR application running on the target user's device. 
This attacker has full access to the APIs provided by the VR device and host system and controls a graphics rendering pipeline that the attacker can use to provide visual stimuli to the target user. In the case of a multiplayer or server-based application, the attacker can process this data arbitrarily before streaming it to a server.\medskip

\noindent \textbf{Privileged Attacker III} (the ``Server Adversary'').\\
Our third privileged attacker represents the entity controlling the external server used to facilitate multiplayer functionality for the application running on the target user's device. This entity may, in practice, be the same party developing the client-side application (in the case of a ``public server") or an entirely separate entity (a ``private server"). Thus, the same entity often controls Privileged Attackers II and III.
This attacker receives a stream of telemetry data from the client-side application, which it can process arbitrarily before relaying such data to one or more other client devices. 
The inferences available to this attacker are generally weaker than the previous attackers; for example, a client application may receive tracking data at $120$ Hz and broadcast it at $30$ Hz~\cite{vrchat_network}, and audio signals are typically heavily compressed before being broadcast.\medskip

\noindent \textbf{Non-Privileged Attacker} (the ``User Adversary'').\\
A non-privileged attacker represents a second end-user of the same multiplayer application as the target user. This attacker receives low-fidelity telemetry and audio streams from the external server for legitimate purposes, such as rendering an avatar representing the target user. They can also interact with the target user as permitted by the application, such as to provide stimuli and observe the target's response. While the audio and telemetry streams are likely highly processed and filtered by this point, they are typically still sufficient to observe the general behavior of the target user.\medskip

The goal of this paper is to understand the capabilities of an adversarial VR application. As such, we evaluate all threats from the perspective of Privileged Attacker II, while noting which other adversaries may be capable of performing the same attacks.

\subsection{Observable Attribute Classes}
\label{sec:classes}

We now shift our discussion to the broad classes of private user data observable by each of the attackers using only their corresponding data sources. We categorize the collected attributes into primary (captured directly from a data source), secondary (derived deterministically from primary attributes), and inferred (derived from primary and secondary attributes using machine learning).\smallskip

\noindent{\textbf{Spatial Telemetry}.} The first major source of user data is directly from telemetry (namely, the position and orientation of the VR headset and controllers over time). Such data is useful for revealing certain anthropometric measurements, such as height and wingspan. While all attackers can observe telemetry to some extent, less privileged attackers are likely to experience degraded precision when estimating these metrics due to the use of intermediate filtering and processing. For example, we found that privileged attackers I and II can determine a user's interpupillary distance (IPD) from telemetry to within $0.1$mm, but IPD is difficult for privileged attackers III and non-privileged attackers to ascertain.\smallskip

\noindent{\textbf{Device Specifications}.} Another class of attack aims to use VR-specific heuristics to determine information about the VR device and the user's host computer. Of course, privileged attackers I and II can directly query device specifications such as resolution and field of view (FOV) from available system APIs; however, we will later demonstrate how even non-privileged attackers can attempt to learn some of this information by creating puzzles that only users of high-fidelity devices can feasibly solve. Determining the specifications of a user's device can reveal personal information about the users themselves. For instance, the cost of commercially-available VR setups spans at least two orders of magnitude; determining the exact hardware of a target user may reveal their income/wealth. \smallskip

\noindent{\textbf{Network Observations}.} An additional source of information about a target user is the observation of network characteristics. While not necessarily unique to virtual reality, attacks that leverage network observations to geolocate users are a natural fit for virtual telepresence applications, which often facilitate the use of multiple game servers to minimize perceived latency~\cite{vrchat_network}. Thus, privileged attackers II and III can efficiently execute such attacks. \smallskip

\noindent{\textbf{Behavioral Observations}.} Behavioral observations are a fourth key source of private information enabled by virtual reality applications, and observing how users react to carefully chosen stimuli can reveal a wide variety of personal information. Attacks based on observing user behavior typically require less privilege than other types of attacks discussed herein, with even non-privileged attackers typically receiving enough information to observe general user interactions. We also include listening to user vocalizations in this category (audio), although one could also consider it a category.

\subsection{Related Work}

\subsubsection{Motion}

For decades, researchers have shown that individuals reveal information about themselves via their motion. As early as 1977, Cutting and Kozlowski demonstrated that people can identify their friends just by viewing the motion of 8 tracked points affixed to the body \cite{cutting_recognizing_1977}, and that the gender of the participants could be identified by a stranger with statistically significant accuracy \cite{kozlowski_recognizing_1977}.
More recently, Pollick et al. (2005) \cite{pollick_gender_2005} used statistical techniques to achieve accurate identification of gender from motion, and Jain et al. (2016) \cite{jain_is_2016} found that the age of a participant can also be accurately determined by their motion patterns.


\subsubsection{VR Privacy}

Because virtual reality devices generate a stream of data relating to the motion of a user, and people are known to subconsciously reveal information about themselves via their motion, it is natural to question the extent to which personal attributes are inferable in VR. According to popular literature reviews of the field \cite{convergenece_VR_social_media, priv_implications_VR, priv_metaverse, metaverse_privacy_1, metaverse_intro, VR_disclosure_info, VR_review_anony, dick_balancing_2021, de_guzman_security_2020, https://doi.org/10.48550/arxiv.2205.00208, 10.1145/2580723.2580730, https://doi.org/10.48550/arxiv.2301.05940}, the vast majority of existing VR privacy research has focused on passive observation.

One major area of interest for researchers has been uniquely identifying users based on their observed motion patterns in VR. A variety of laboratory studies \cite{pfeuffer_behavioural_2019, pfeuffer_behavioural_2019, liebers_understanding_2021, tricomi_you_2022} have been conducted in which 16 to 511 users were passively observed while performing a number of standardized tasks in VR, after which machine learning techniques were used to uniquely identify them with accuracies ranging from 40\% to 95\%. More recently, Nair et al. \cite{https://doi.org/10.48550/arxiv.2302.08927} passively observed 55,541 players of a popular VR game, and were later able to uniquely identify players with 94\% accuracy based on their motion.

A second line of work has focused on identifying specific demographic attributes like age and gender based on VR recordings \cite{9417659, 10.1145/3491102.3502008, 8990850}. As with the identification studies, these works utilize the passive observation of people using standard VR applications.

While the notion of malicious VR application design has been qualitatively discussed since 2020 \cite{9319051}, no user study has yet been produced that directly implements and evaluates an adversarially-designed VR application. However, the success of gamified data harvesting on traditional social media platforms suggests that even stronger capabilities may be available to an active attacker. 


\subsubsection{Gamified Data Harvesting}

In 2018, the British political consulting firm Cambridge Analytica was revealed to be in possession of personal data from up to 87 million Facebook users.
Subsequent analysis revealed that most of this data was collected through Facebook quizzes designed to seem like fun personality assessments while actually building a detailed profile of user data \cite{isaak2018user, schneble2018cambridge, berghel2018malice}.

Gamified data collection mechanisms bypass the normal cognitive filters associated with data privacy by taking advantage of users' innate desire to perform optimally when completing challenges. Presenting a data-revealing question as a puzzle or element serving a legitimate role in a broader game has proven effective at obscuring the hidden intent to collect personal information  \cite{gamified, noauthor_hr_2018}.

\subsection{Motivation}

In this paper, we seek to combine the known possibility of inferring private data attributes from VR motion data with the success of gamified data harvesting in conventional social platforms to explore data harvesting attacks made possible by adversarial game design in VR. Our interest in this field is motivated by two simple observations.
First, gaming is the predominant driver of VR adoption today \cite{steam_most_used}, providing ample opportunity to disguise data collection mechanisms as VR game elements.
Simultaneously, the social platforms exploited by Cambridge Analytica are now dominant players in the AR/VR space. It is natural to assume that some may wish to use the same techniques that have proven successful on conventional social media platforms in the data-rich environment of VR.




\section{VR Privacy Attacks}
\label{sec:Privacy_attacks}

Next, we describe specific examples of adversarial game elements designed to extract VR user data, corresponding to the broad observable attribute classes detailed in \S\ref{sec:classes}. The goal of this section is not to be exhaustive with respect to the wide variety of interactive elements that can reveal user information, but rather to exemplify strategies for collecting various types of attributes using specific mechanisms that we later evaluate in our user study (\S\ref{sec:Experimental_Design}).

\subsection{Biometrics}
\label{sec:biom}

\vspace{-1em}

\begin{figure}[!h]
    \centering
    \includegraphics[width=0.75 \linewidth]{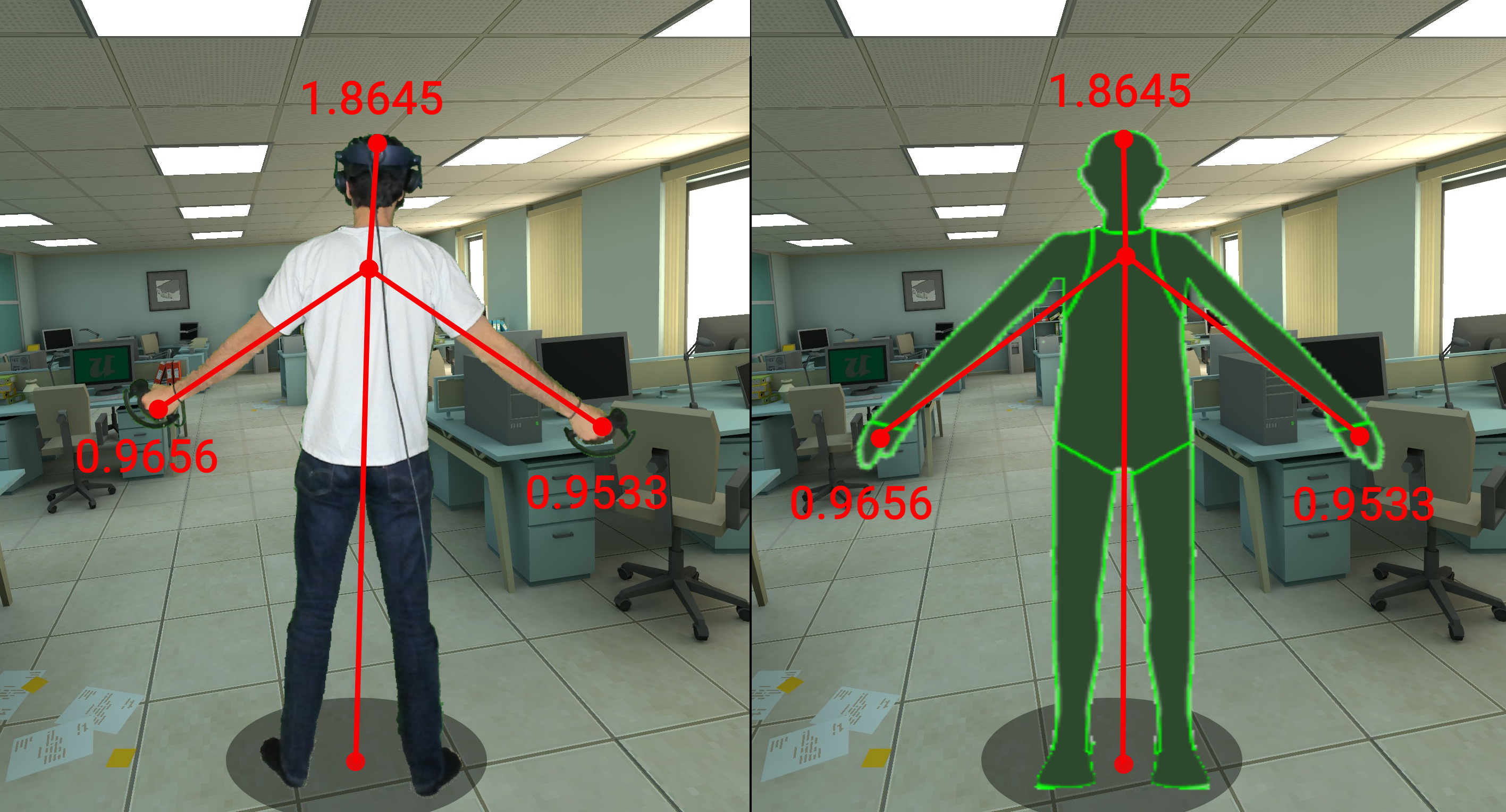}
    \vspace{-1em}
    \caption{Measuring user anthropometrics from telemetry.}
    \label{fig:geometry}
\end{figure}

\vspace{-1em}

\begin{figure}[!h]
    \centering
    \includegraphics[width=0.75 \linewidth]{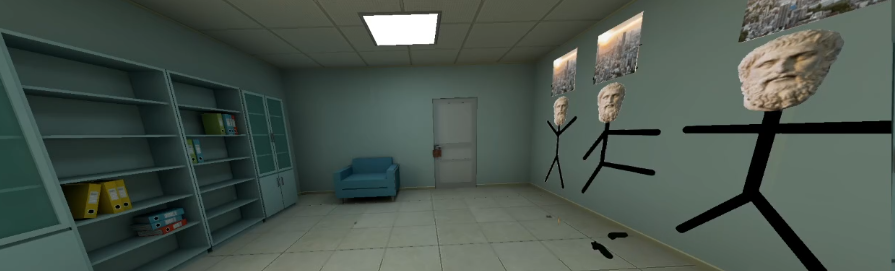}
    \vspace{-1em}
    \caption{Adversarial measurement of wingspan.}
    \label{fig:wingspanrm}
\end{figure}

\noindent{\textbf{Continuous Anthropometrics}.} 
Basic anthropometrics provide a simple yet compelling example of the dangers of adversarial design.
Fig.~\ref{fig:geometry} illustrates how attackers can passively measure a user's height and wingspan from VR telemetry.
However, users are unlikely to naturally stand in a position that readily facilitates the measurement of wingspan. Therefore, Fig.~\ref{fig:wingspanrm} depicts a pose-based game element designed to subtly induce a standing position more conducive to precise anthropometric measurement.

\vspace{-1em}

\vspace{-1em}

\begin{figure}[!h]
    \centering
    \includegraphics[width=0.75 \linewidth]{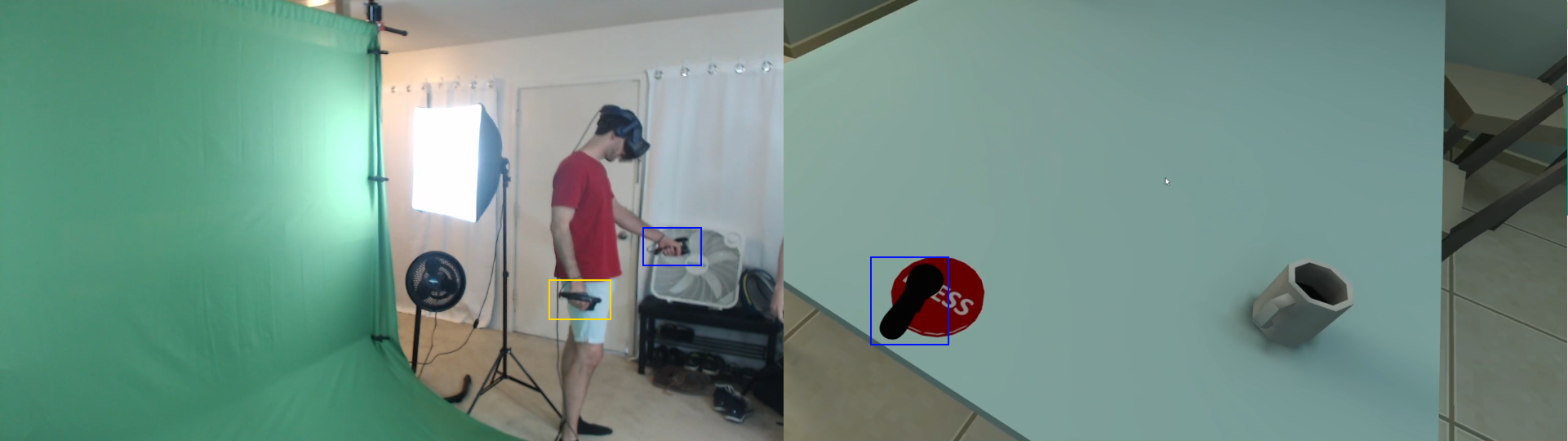}
    \vspace{-1em}
    \caption{Estimating handedness from behavior.}
    \label{fig:handedness}
\end{figure}

\noindent{\textbf{Binary Anthropometrics}.}
An attacker can collect binary anthropometrics, which include characteristics such as longer-arm and dominant handedness, both directly from telemetry (e.g., ``which hand moves more?") and from behavior (e.g., ``which hand is used to press a button?"). Fig.~\ref{fig:handedness} illustrates an example process of determining a user's handedness by including a small button that requires precise manipulation, suggesting the use of one's dominant hand.

\vspace{-1em}

\begin{figure}[H]
    \centering
    \includegraphics[width=\linewidth]{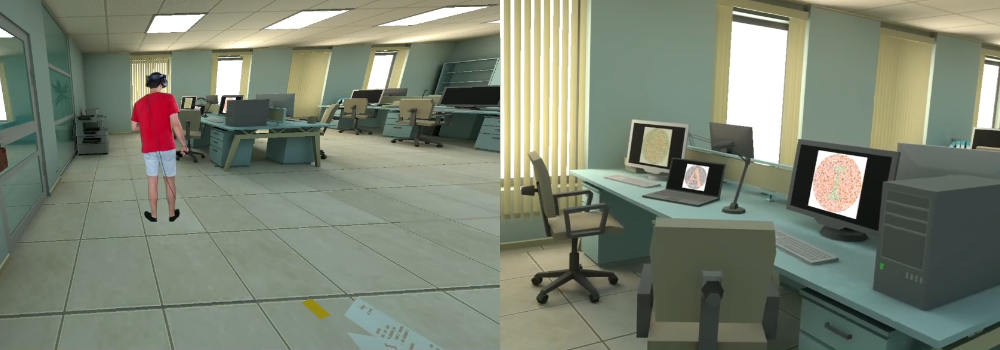}
    \caption{VR puzzle revealing deuteranopia.}
    \label{fig:color}
\end{figure}

\noindent{\textbf{Vision}.} 
VR attackers can carefully construct interactive elements that secretly reveal aspects of a player's visual acuity, such as nearsightedness, farsightedness, or color blindness. For example, Fig.~\ref{fig:color} shows a puzzle element of a VR game that appears innocuous to most users but is not solvable by users with red-green color blindness (deuteranopia), thus revealing the presence of that condition.

\begin{figure}[!h]
    \centering
    \begin{subfigure}{.41\linewidth}
        \centering
        \includegraphics[width=\linewidth]{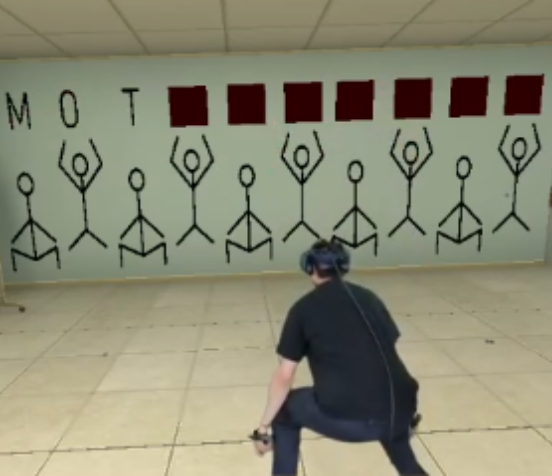}
    \end{subfigure}%
    \hfill
    \begin{subfigure}{.55 \linewidth}
        \centering
        \includegraphics[width=\linewidth]{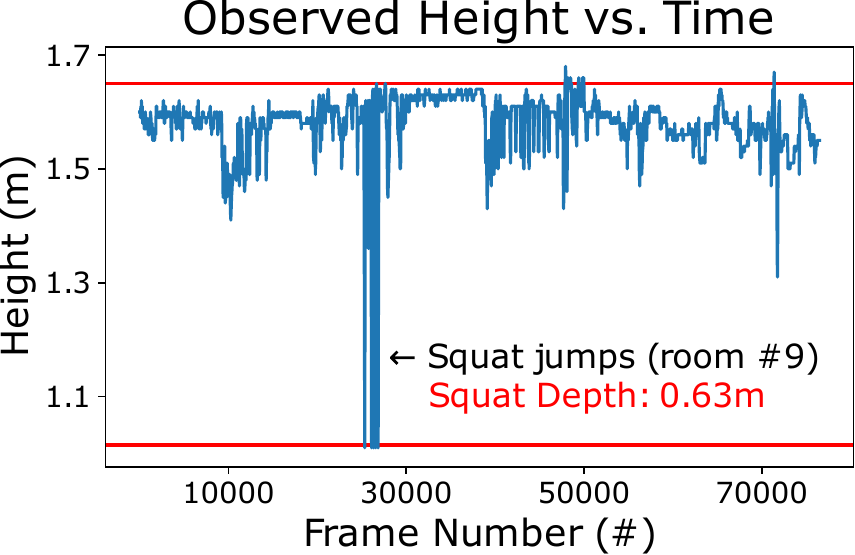}
    \end{subfigure}%
    \caption{Measurement of physical fitness.}
    \label{fig:fitness}
\end{figure}

\noindent{\textbf{Fitness}.} Attackers can also use behavioral and telemetric measurements to asses a subject's degree of physical fitness. Fig.~\ref{fig:fitness} illustrates a virtual room designed to elicit physical activity and shows the resulting metric of physical fitness measurable on a headset position (y-coordinate) vs. time graph. We observed that a squat depth of less than $25\%$ of height corresponded to low physical fitness, though other metrics can also be used. An extreme lack of fitness may reveal a participant's age or the presence of physical disabilities.

\begin{figure}[!h]
    \centering
    \includegraphics[width=\linewidth]{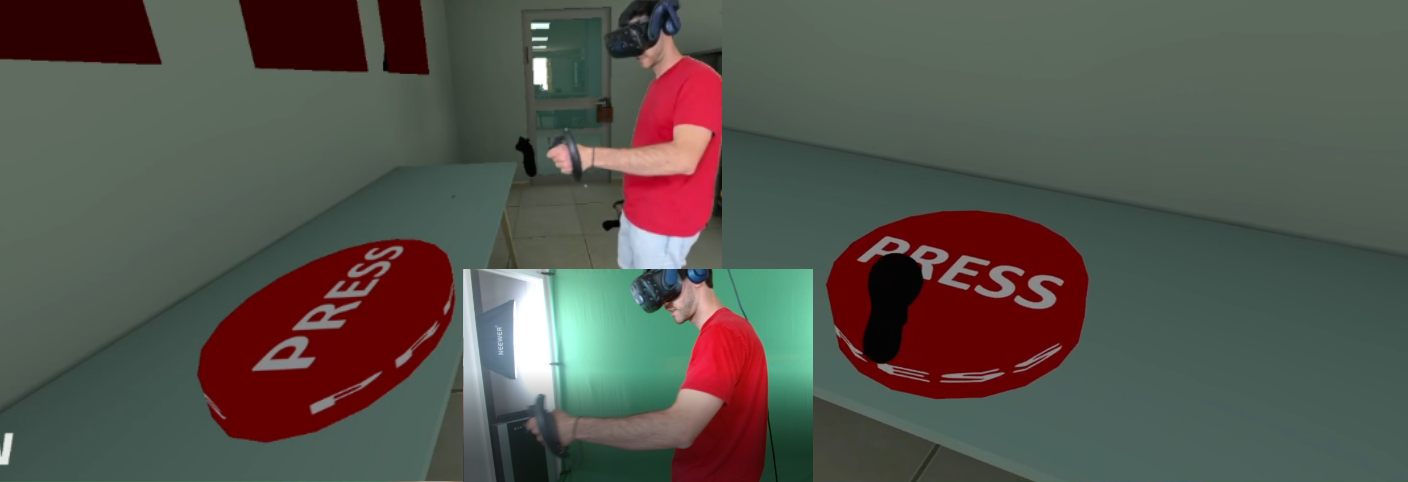}
    \caption{VR puzzle measuring reaction time.}
    \label{fig:reaction}
\end{figure}

\noindent{\textbf{Reaction Time}.} Fig.~\ref{fig:reaction} shows a VR environment adversarially constructed to reveal the participant's reaction time by measuring the time interval between a visual stimulus and motor response. Reaction time is strongly correlated with age \cite{woods_age-related_2015}.

\subsection{Environment}
\begin{figure}[H]
    \centering
    \includegraphics[width=0.8 \linewidth]{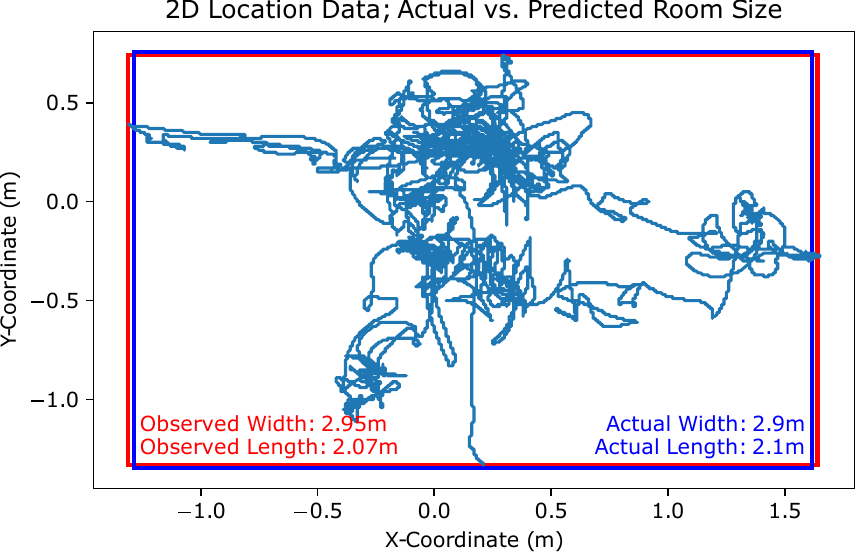}
    \caption{Estimating room size from spatial telemetry.}
    \label{fig:room}
\end{figure}

\noindent{\textbf{Room Size}.} Fig.~\ref{fig:room} shows how an attacker could estimate the size of a user's physical environment by tracking their virtual movements. 
Virtual environments can be designed to contain interactive elements which specifically encourage the participant to explore the boundaries of their physical environment.

\begin{figure}[!h]
    \centering
    \includegraphics[width=0.8 \linewidth]{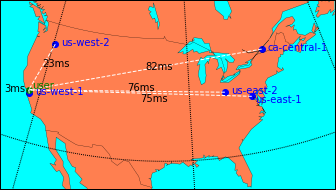}
    \caption{Estimating user location from network latency.}
    \label{fig:network}
\end{figure}

\noindent{\textbf{Geolocation}.} 
Fig.~\ref{fig:network} shows how observing the round-trip delay between a client device and multiple game servers (proximity) can reveal an end user's location (locality) via multilateration. 
A non-privileged attacker could use the round trip delay of audio signals as an approximate measure of latency. 

\subsection{Device Specifications}
\begin{figure}[H]
    \centering
    \begin{subfigure}{.49\linewidth}
        \centering
        \includegraphics[width=\linewidth]{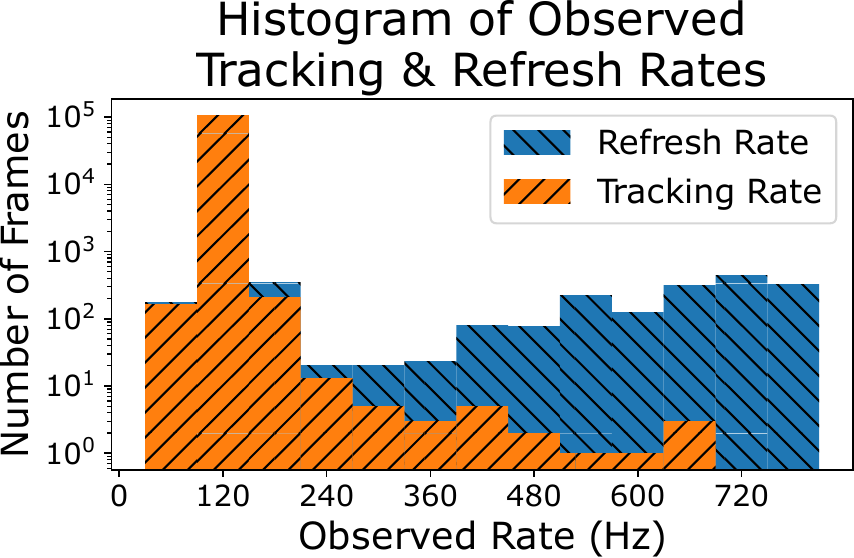}
        \caption{Measuring headset and tracking refresh rate from device API throughput (log scale).}
        \label{fig:histogram}
    \end{subfigure}%
    \hfill
    \begin{subfigure}{.47\linewidth}
        \centering
        \includegraphics[width=\linewidth]{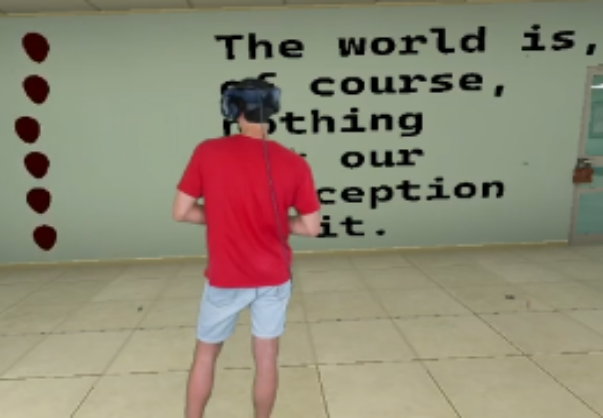}
        \caption{Environment designed to reveal headset refresh rate via perceived motion differences.}
        \label{fig:refresh}
    \end{subfigure}%
    \caption{Methods of attaining VR device metrics.}
\end{figure}

\noindent{\textbf{VR Device}.} 
We assume that privileged attackers I and II have intrinsic knowledge of the VR device specifications via direct API interaction. 
Fig.~\ref{fig:histogram} shows how privileged attackers may use the observed update frequency of telemetry data to determine the polling rate of a target user's controller tracking.
Further, Fig.~\ref{fig:refresh} shows how even a non-privileged attacker can construct a virtual environment that replicates the ``UFO test'' \cite{noauthor_blur_nodate}, which users perceive differently depending on their devices' refresh rate (see puzzle $15$ in Appendix~\ref{sec:VR_puzzles}). 
Currently, determining refresh rate, resolution, and field of view is sufficient to reveal the exact model of the VR device.\medskip

\noindent{\textbf{Host Device}.} Privileged attackers can also embed a variety of standardized benchmarks in their source code to assess the quality of the target user's host device (gaming computer). An attacker can use metrics such as CPU power, GPU power, and network bandwidth to reveal the age and price tier of the system and, thus, potentially correlate the spending power of the target user.

\begin{figure}[!h]
    \centering
    \begin{subfigure}{.48\linewidth}
        \centering
        \includegraphics[width=\linewidth]{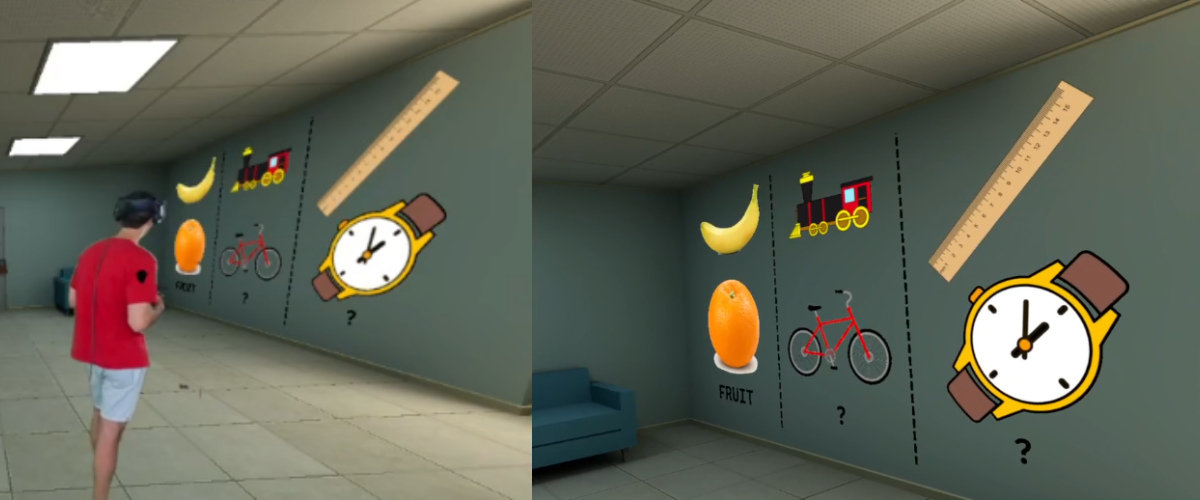}
        \caption{abstraction}
        \label{fig:abstraction}
    \end{subfigure}%
    \hfill
    \begin{subfigure}{.48\linewidth}
        \centering
        \includegraphics[width=\linewidth]{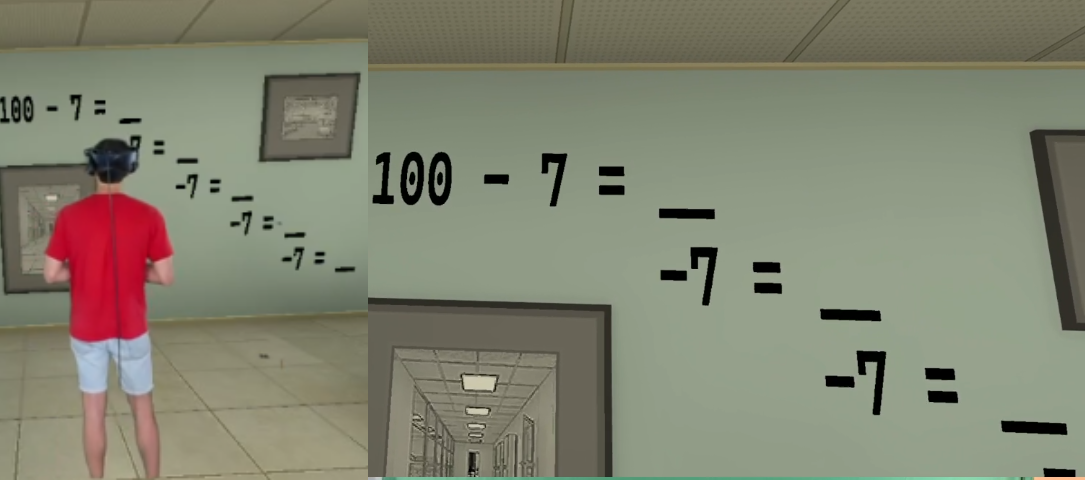}
        \caption{attention}
        \label{fig:attention}
    \end{subfigure}%
    \vskip\baselineskip
    \begin{subfigure}{.48\linewidth}
        \centering
        \includegraphics[width=\linewidth]{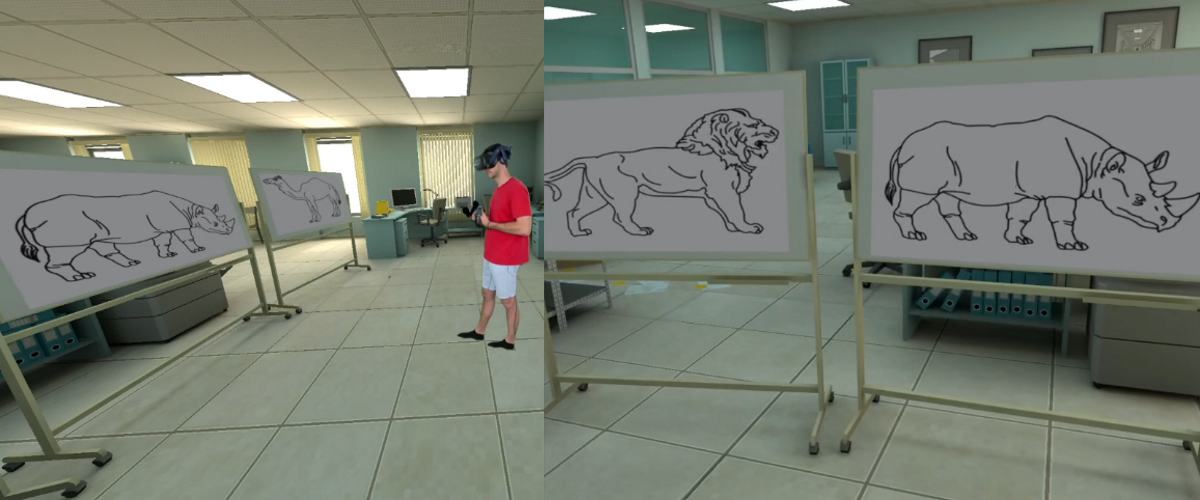}
        \caption{naming}
        \label{fig:naming}
    \end{subfigure}%
    \hfill
    \begin{subfigure}{.48\linewidth}
        \centering
        \includegraphics[width=\linewidth]{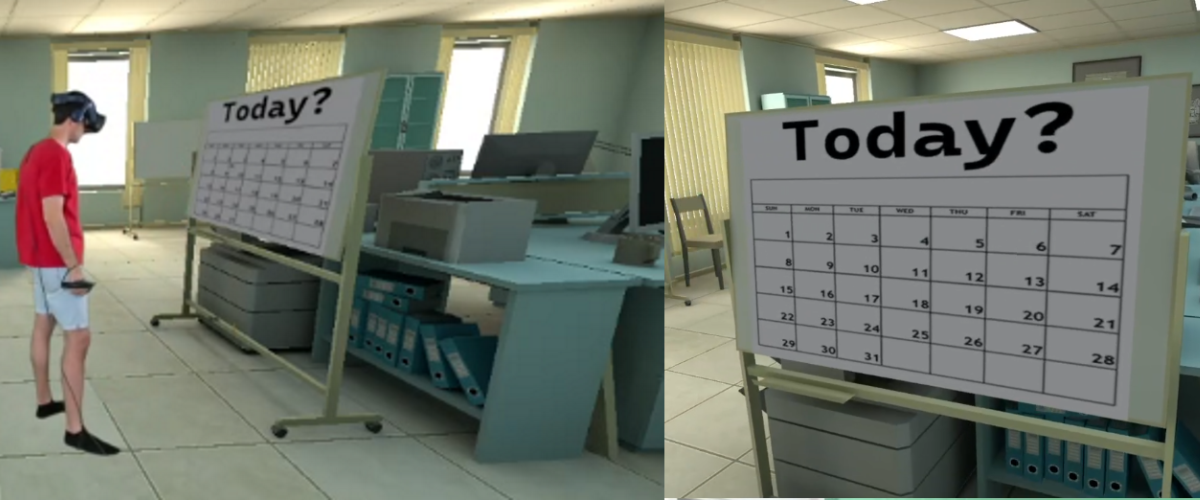}
        \caption{orientation}
        \label{fig:orientation}
    \end{subfigure}%
    \caption{Methods of measuring cognitive acuity.}
    \label{fig:MOCA}
\end{figure}

\subsection{Acuity (MoCA)}
A number of standardized cognitive, diagnostic, and aptitude tests can be adapted for (and hidden within) VR environments. 
Fig.~\ref{fig:MOCA} illustrates VR environments designed to covertly asses four categories of the Montreal Cognitive Assessment (MoCA): abstraction (\ref{fig:abstraction}), attention (\ref{fig:attention}), naming (\ref{fig:naming}), and orientation (\ref{fig:orientation}).

\begin{table*}
\setlength{\tabcolsep}{8.5pt}
\resizebox{\textwidth}{!}{%
\begin{tabular}{l|ccc|ccc|ccc|c|c|c|}
\cline{2-13}
 & \multicolumn{3}{c|}{\textbf{Head}} & \multicolumn{3}{c|}{\textbf{Left}} & \multicolumn{3}{c|}{\textbf{Right}} & \multirow{2}{*}{\textbf{Device}} & \multirow{2}{*}{\textbf{Microphone}} & \multirow{2}{*}{\textbf{Behavior}} \\ \cline{2-10}
 & \multicolumn{1}{c|}{\textbf{X}} & \multicolumn{1}{c|}{\textbf{Y}} & \textbf{Z} & \multicolumn{1}{c|}{\textbf{X}} & \multicolumn{1}{c|}{\textbf{Y}} & \textbf{Z} & \multicolumn{1}{c|}{\textbf{X}} & \multicolumn{1}{c|}{\textbf{Y}} & \textbf{Z} &  &  &  \\ \hline
\multicolumn{1}{|l|}{\textbf{Height}} & \multicolumn{1}{c|}{} & \multicolumn{1}{c|}{\Checkmark} &  & \multicolumn{1}{c|}{} & \multicolumn{1}{c|}{} &  & \multicolumn{1}{c|}{} & \multicolumn{1}{c|}{} &  &  &  &  \\ \hline
\multicolumn{1}{|l|}{\textbf{Left Arm}} & \multicolumn{1}{c|}{\Checkmark} & \multicolumn{1}{c|}{} & \Checkmark & \multicolumn{1}{c|}{\Checkmark} & \multicolumn{1}{c|}{} & \Checkmark & \multicolumn{1}{c|}{} & \multicolumn{1}{c|}{} &  &  &  &  \\ \hline
\multicolumn{1}{|l|}{\textbf{Right Arm}} & \multicolumn{1}{c|}{\Checkmark} & \multicolumn{1}{c|}{} & \Checkmark & \multicolumn{1}{c|}{} & \multicolumn{1}{c|}{} &  & \multicolumn{1}{c|}{\Checkmark} & \multicolumn{1}{c|}{} & \Checkmark &  &  &  \\ \hline
\multicolumn{1}{|l|}{\textbf{Longer Arm}} & \multicolumn{1}{c|}{\Checkmark} & \multicolumn{1}{c|}{} & \Checkmark & \multicolumn{1}{c|}{\Checkmark} & \multicolumn{1}{c|}{} & \Checkmark & \multicolumn{1}{c|}{\Checkmark} & \multicolumn{1}{c|}{} & \Checkmark &  &  &  \\ \hline
\multicolumn{1}{|l|}{\textbf{Handedness}} & \multicolumn{1}{c|}{} & \multicolumn{1}{c|}{} &  & \multicolumn{1}{c|}{\Checkmark} & \multicolumn{1}{c|}{\Checkmark} & \Checkmark & \multicolumn{1}{c|}{\Checkmark} & \multicolumn{1}{c|}{\Checkmark} & \Checkmark &  &  &  \\ \hline
\multicolumn{1}{|l|}{\textbf{Wingspan}} & \multicolumn{1}{c|}{} & \multicolumn{1}{c|}{} &  & \multicolumn{1}{c|}{\Checkmark} & \multicolumn{1}{c|}{} & \Checkmark & \multicolumn{1}{c|}{\Checkmark} & \multicolumn{1}{c|}{} & \Checkmark &  &  &  \\ \hline
\multicolumn{1}{|l|}{\textbf{Room Length}} & \multicolumn{1}{c|}{\Checkmark} & \multicolumn{1}{c|}{} &  & \multicolumn{1}{c|}{} & \multicolumn{1}{c|}{} &  & \multicolumn{1}{c|}{} & \multicolumn{1}{c|}{} &  &  &  &  \\ \hline
\multicolumn{1}{|l|}{\textbf{Room Width}} & \multicolumn{1}{c|}{} & \multicolumn{1}{c|}{} & \Checkmark & \multicolumn{1}{c|}{} & \multicolumn{1}{c|}{} &  & \multicolumn{1}{c|}{} & \multicolumn{1}{c|}{} &  &  &  &  \\ \hline
\multicolumn{1}{|l|}{\textbf{Room Size}} & \multicolumn{1}{c|}{\Checkmark} & \multicolumn{1}{c|}{} & \Checkmark & \multicolumn{1}{c|}{} & \multicolumn{1}{c|}{} &  & \multicolumn{1}{c|}{} & \multicolumn{1}{c|}{} &  &  &  &  \\ \hline
\multicolumn{1}{|l|}{\textbf{IPD}} & \multicolumn{1}{c|}{} & \multicolumn{1}{c|}{} &  & \multicolumn{1}{c|}{} & \multicolumn{1}{c|}{} &  & \multicolumn{1}{c|}{} & \multicolumn{1}{c|}{} &  & \Checkmark &  &  \\ \hline
\multicolumn{1}{|l|}{\textbf{Eyesight}} & \multicolumn{1}{c|}{} & \multicolumn{1}{c|}{} &  & \multicolumn{1}{c|}{} & \multicolumn{1}{c|}{} &  & \multicolumn{1}{c|}{} & \multicolumn{1}{c|}{} &  &  &  & \Checkmark \\ \hline
\multicolumn{1}{|l|}{\textbf{Color Blindness}} & \multicolumn{1}{c|}{} & \multicolumn{1}{c|}{} &  & \multicolumn{1}{c|}{} & \multicolumn{1}{c|}{} &  & \multicolumn{1}{c|}{} & \multicolumn{1}{c|}{} &  &  &  & \Checkmark \\ \hline
\multicolumn{1}{|l|}{\textbf{Locality}} & \multicolumn{1}{c|}{} & \multicolumn{1}{c|}{} &  & \multicolumn{1}{c|}{} & \multicolumn{1}{c|}{} &  & \multicolumn{1}{c|}{} & \multicolumn{1}{c|}{} &  &  & \Checkmark &  \\ \hline
\multicolumn{1}{|l|}{\textbf{Device Refresh Rate}} & \multicolumn{1}{c|}{} & \multicolumn{1}{c|}{} &  & \multicolumn{1}{c|}{} & \multicolumn{1}{c|}{} &  & \multicolumn{1}{c|}{} & \multicolumn{1}{c|}{} &  &  &  & \Checkmark \\ \hline
\multicolumn{1}{|l|}{\textbf{Tracking Refresh Rate}} & \multicolumn{1}{c|}{} & \multicolumn{1}{c|}{} &  & \multicolumn{1}{c|}{} & \multicolumn{1}{c|}{} &  & \multicolumn{1}{c|}{} & \multicolumn{1}{c|}{} &  & \Checkmark &  &  \\ \hline
\multicolumn{1}{|l|}{\textbf{Device Resolution}} & \multicolumn{1}{c|}{} & \multicolumn{1}{c|}{} &  & \multicolumn{1}{c|}{} & \multicolumn{1}{c|}{} &  & \multicolumn{1}{c|}{} & \multicolumn{1}{c|}{} &  & \Checkmark &  &  \\ \hline
\multicolumn{1}{|l|}{\textbf{Device FOV}} & \multicolumn{1}{c|}{} & \multicolumn{1}{c|}{} &  & \multicolumn{1}{c|}{} & \multicolumn{1}{c|}{} &  & \multicolumn{1}{c|}{} & \multicolumn{1}{c|}{} &  & \Checkmark &  &  \\ \hline
\multicolumn{1}{|l|}{\textbf{VR Device}} & \multicolumn{1}{c|}{} & \multicolumn{1}{c|}{} &  & \multicolumn{1}{c|}{} & \multicolumn{1}{c|}{} &  & \multicolumn{1}{c|}{} & \multicolumn{1}{c|}{} &  & \Checkmark &  & \Checkmark \\ \hline
\multicolumn{1}{|l|}{\textbf{Computing Power}} & \multicolumn{1}{c|}{} & \multicolumn{1}{c|}{} &  & \multicolumn{1}{c|}{} & \multicolumn{1}{c|}{} &  & \multicolumn{1}{c|}{} & \multicolumn{1}{c|}{} &  & \Checkmark &  &  \\ \hline
\multicolumn{1}{|l|}{\textbf{Languages}} & \multicolumn{1}{c|}{} & \multicolumn{1}{c|}{} &  & \multicolumn{1}{c|}{} & \multicolumn{1}{c|}{} &  & \multicolumn{1}{c|}{} & \multicolumn{1}{c|}{} &  &  &  & \Checkmark \\ \hline
\multicolumn{1}{|l|}{\textbf{Physical Fitness}} & \multicolumn{1}{c|}{} & \multicolumn{1}{c|}{\Checkmark} &  & \multicolumn{1}{c|}{} & \multicolumn{1}{c|}{} &  & \multicolumn{1}{c|}{} & \multicolumn{1}{c|}{} &  &  &  &  \\ \hline
\multicolumn{1}{|l|}{\textbf{Reaction Time}} & \multicolumn{1}{c|}{} & \multicolumn{1}{c|}{} &  & \multicolumn{1}{c|}{} & \multicolumn{1}{c|}{} &  & \multicolumn{1}{c|}{} & \multicolumn{1}{c|}{} &  &  &  & \Checkmark \\ \hline
\multicolumn{1}{|l|}{\textbf{MOCA}} & \multicolumn{1}{c|}{} & \multicolumn{1}{c|}{} &  & \multicolumn{1}{c|}{} & \multicolumn{1}{c|}{} &  & \multicolumn{1}{c|}{} & \multicolumn{1}{c|}{} &  &  &  & \Checkmark \\ \hline
\multicolumn{1}{|l|}{\textbf{Gender}} & \multicolumn{1}{c|}{} & \multicolumn{1}{c|}{\Checkmark} &  & \multicolumn{1}{c|}{\Checkmark} & \multicolumn{1}{c|}{} & \Checkmark & \multicolumn{1}{c|}{\Checkmark} & \multicolumn{1}{c|}{} & \Checkmark & \Checkmark & \Checkmark &  \\ \hline
\multicolumn{1}{|l|}{\textbf{Age}} & \multicolumn{1}{c|}{} & \multicolumn{1}{c|}{\Checkmark} &  & \multicolumn{1}{c|}{} & \multicolumn{1}{c|}{} &  & \multicolumn{1}{c|}{} & \multicolumn{1}{c|}{} &  &  &  & \Checkmark \\ \hline
\multicolumn{1}{|l|}{\textbf{Ethnicity}} & \multicolumn{1}{c|}{} & \multicolumn{1}{c|}{\Checkmark} &  & \multicolumn{1}{c|}{} & \multicolumn{1}{c|}{} &  & \multicolumn{1}{c|}{} & \multicolumn{1}{c|}{} &  &  & \Checkmark & \Checkmark \\ \hline
\multicolumn{1}{|l|}{\textbf{Disability Status (Mental)}} & \multicolumn{1}{c|}{} & \multicolumn{1}{c|}{} &  & \multicolumn{1}{c|}{} & \multicolumn{1}{c|}{} &  & \multicolumn{1}{c|}{} & \multicolumn{1}{c|}{} &  &  &  & \Checkmark \\ \hline
\multicolumn{1}{|l|}{\textbf{Disability Status (Physical)}} & \multicolumn{1}{c|}{} & \multicolumn{1}{c|}{\Checkmark} &  & \multicolumn{1}{c|}{} & \multicolumn{1}{c|}{} &  & \multicolumn{1}{c|}{} & \multicolumn{1}{c|}{} &  &  &  &  \\ \hline
\multicolumn{1}{|l|}{\textbf{Identity}} & \multicolumn{1}{c|}{\Checkmark} & \multicolumn{1}{c|}{\Checkmark} & \Checkmark & \multicolumn{1}{c|}{\Checkmark} & \multicolumn{1}{c|}{\Checkmark} & \Checkmark & \multicolumn{1}{c|}{\Checkmark} & \multicolumn{1}{c|}{\Checkmark} & \Checkmark & \Checkmark & \Checkmark & \Checkmark \\ \hline
\end{tabular}%
}
\caption{VR device sensors associated with each attack.}
\label{tab:sensors}
\end{table*}

\subsection{Demographics}
\begin{figure}[!h]
    \centering
    \includegraphics[width=\linewidth]{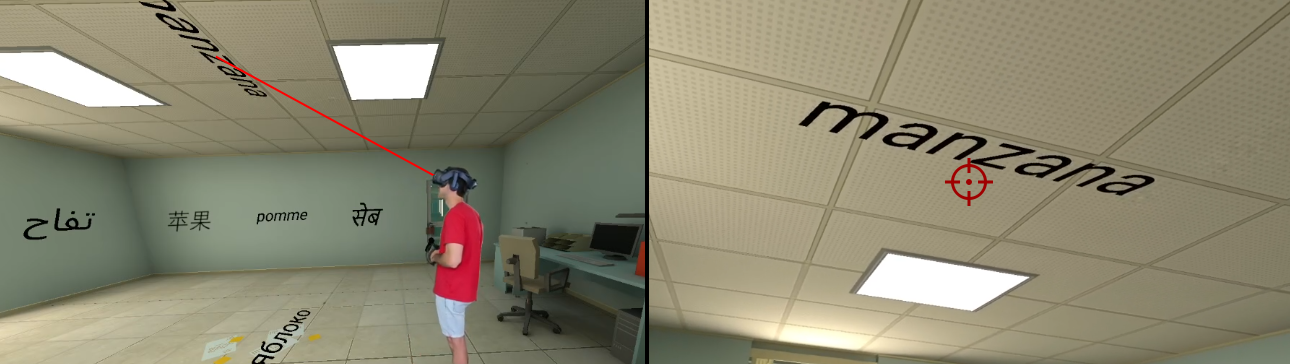}
    \caption{Determining language from user behavior.}
    \label{fig:language}
\end{figure}

\noindent{\textbf{Language}.} There are a number of ways to ascertain a user's spoken language(s) in VR, including via speech recognition. Fig.~\ref{fig:language} illustrates how a non-privileged attacker can observe a user's direction of gaze while solving a puzzle to reveal the languages they speak.\medskip

\noindent{\textbf{Vocal Characteristics}.} Listening to the voice of a user may reveal key demographic attributes such as age, gender, and ethnicity~\cite{becker_primaryobjectsvoice-gender_2022, garg_speech-accent-recognition_2022}. Shared VR environments with voice streaming provide a strong opportunity to exploit voice analysis, as attackers can cue target users to speak certain words or phrases that reveal more information, such as by requiring players to speak passwords aloud.\medskip

\noindent{\textbf{Inferred Attributes}.} While most demographic attributes cannot be observed directly from VR data, attackers can often accurately infer them from primary data attributes. For example, height, wingspan, and IPD correlate strongly with gender, while eyesight, reaction time, and fitness correlate with age.
While not possible to measure accurately in this study, we also suggest that in practice, information about room size, VR device type, and computing power could be used together to infer the income or wealth of a user.\medskip

\noindent{\textbf{Identity}.}
Finally, the various attributes discussed above can be combined to uniquely identify VR users. While previous work has already demonstrated the deanonymization of VR users \cite{miller_personal_2020} within small groups, the number of data attributes presented here far exceed the $15$ or so thought necessary to uniquely identify every individual in the United States \cite{rocher}.

\begin{figure*}[!ht]
    \centering
    \includegraphics[width= \linewidth]{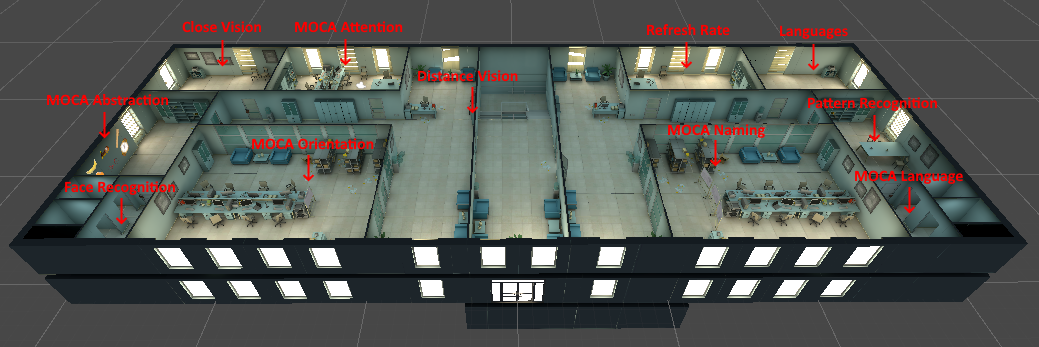}
    \caption{Virtual office building of the ``MetaData'' game, hosting the puzzle rooms.}
    \label{fig:virtual_building}
\end{figure*}

\subsection{Summary}
Table \ref{tab:sensors} summarizes the VR privacy attacks presented in this section along with the VR device sensors or sources of information associated with each attack. 
The incredible volume of information exposed by a metaverse user, with at least 18 telemetry values collected 60 times per second or more, provides a vast amount of data from which adversarial inferences can be made.
In all, we have identified dozens of unique data attributes, ranging from biometrics and demographics to behavioral and environmental measurements, that can be observed from users in VR via adversarial game design.

Of course, these attacks are by no means exhaustive, with many further attributes likely being observable that we have not discussed.
Instead, our examples serve to illustrate the wide scope of observations available to VR adversaries and the ability to capture a comprehensive user attribute profile that would otherwise have involved aggregating data across several different devices.

Having discussed in great detail the theoretical information flow, adversaries, and plausible inferences of metaverse environments, the remainder of this paper focuses on the experimental validation and quantification of these threats.

While a detailed description of the evaluated attacks is necessary for the completeness of this study, it is not our intention to focus on any particular attributes. Rather, our goal, as highlighted by the experimental design described below, is to generally demonstrate the extent to which adversarial game design can enable the collection of sensitive data attributes in VR.






\section{Experimental Design}
\label{sec:Experimental_Design}

In this section, we describe ``MetaData,'' a virtual reality ``escape room'' game designed as a case study for understanding how adversarial game design enhances attacker capabilities in virtual reality. 
The question we aim to answer is whether, and to what degree, an attacker can use data collected from consumer-grade VR devices to accurately extract and infer users' private information when aided by the capability to adversarially construct the virtual world and application rather than merely relying on passive observation.

This section details the experimental design, technical setup, and protocol used to answer this important question.
After identifying the privacy-sensitive variables we believed to be accessible within VR (as detailed in \S\ref{sec:Privacy_attacks}), we implemented systematic methods to collect and analyze these variables from within VR applications.

To test the efficacy of these attacks, we designed an "escape room"-style VR game themed as an office building (see Fig.~\ref{fig:virtual_building}). We then disguised the attacks as a set of puzzles within the game, which users were highly motivated to solve to the best of their ability in order to unlock a sequence of doors and win the game.
We describe and illustrate the exact puzzles in detail in Appendix~\ref{sec:VR_puzzles}.

We endeavored to design the experiment such that it did not bluntly reveal the ulterior goal, thereby illustrating how other VR applications could also accomplish the same goal covertly. To this end, we also added innocuous (i.e., ``noisy'') rooms which did not necessarily collect meaningful personal information, but instead served to camouflage the data-harvesting puzzles. 


\begin{figure}[H]
    \centering
    \includegraphics[width=\linewidth]{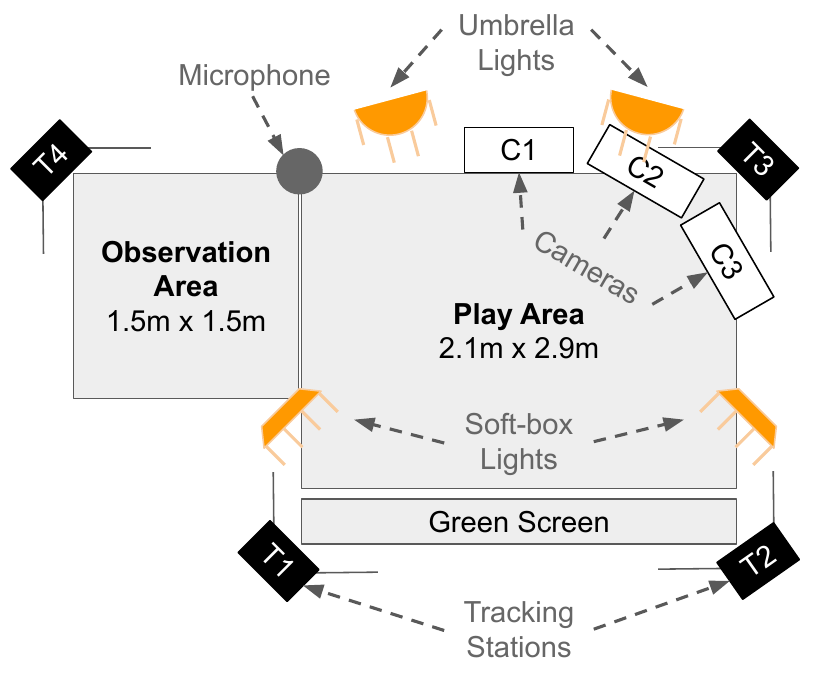}
    \caption{VR laboratory room layout.}
    \label{fig:layout}
\end{figure}

\subsection{Setup and Protocol}
\label{subsec:setup}

\begin{figure}[!htbp]
    \centering
    \includegraphics[width=0.8 \linewidth]{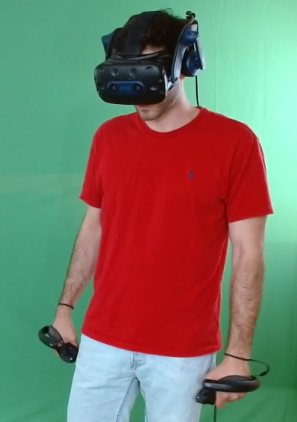}
    \caption{Experimental setup (not a real participant).}
    \label{fig:experimental_setup}
\end{figure}

We recruited $50$ individuals for the experiments (participant distribution given in Appendix~\ref{app:distr}).
After completing a thorough informed consent and orientation process, 
we helped the participants don a VR headset (HTC Vive, Vive Pro 2, or Oculus Quest 2) and its hand-held controllers (Vive Controllers, Valve Index Controllers, or Oculus Quest Controllers, respectively), after which the participant proceeded to play the VR game (see the laboratory room layout in Fig.~\ref{fig:layout} and the primary VR setup in Fig.~\ref{fig:experimental_setup}).
Finally, the participants completed a post-game survey to collect the ``ground truth'' values for attributes of interest. The methods for collecting the true attribute values are summarized in Appendix~\ref{app:instr}.

We tested three devices to determine if there were any noteworthy differences in the findings, which we did not observe other than in IPD (see \S\ref{sec:Biometrics}), and to provide distinct classes for device identification.
Each headset was paired with a gaming computer sufficiently powerful to run it at full fidelity; the main experimental setup had $64$~GB of RAM, an AMD~Ryzen~9~5950X CPU, and an Nvidia RTX~3090 GPU. 
To produce accurate results for room size and geolocation, we also conducted our experiment across four geographically distinct laboratories.

Each experiment lasted approximately $10$--$20$ minutes within VR, plus around $10$ minutes for completing the survey.
Throughout the experiments, we minimized the interactions with the participants and ensured their safety by intervening when they approached a wall in the room.
The experiments remained the same for all participants; we did not alter the game play-through or logic.
The game collected the targeted data points in CSV format during the play-through.
Furthermore, the researchers manually annotated data points for data collection that required game development beyond what is reasonable for this study, e.g., automating voice recognition to register the escape room ``passwords'' (solutions) the participants articulated aloud.
The researchers pressed keys on a keyboard to trigger animations in the virtual environment and teleport the player between rooms. These elements could be automated in a production-ready VR game.

Once the experiment ended, the participants filled out a form with their ground truth, which we used to validate the accuracy of the proposed privacy attacks.
To collect the ground truth unknown to the participants themselves, we performed onsite measurements, e.g., we annotated the VR device and VR-room area, tested their reaction time with a desktop app, and measured their height and wingspan with a metric tape.
Furthermore, knowing that researchers have studied the use of cognitive assessments in the diagnosis of attention disorders~\cite{attention_disorder_VR}, autism~\cite{autism_VR}, PTSD~\cite{ptsd_VR}, and dementia~\cite{dementia_VR}, we chose the Montreal cognitive assessment (MoCA)~\cite{larner_montreal_2017} as a simple example of what advanced, immersive VR games could hide in their play-throughs.
We randomized the order of the VR experiment and paper MoCA test (with half the participants taking the MoCA before and with the other half after the experiment) to neutralize potential biases in either direction.
The exact method of collecting ``ground truth'' measurements for each attribute value is described in Appendix~\ref{app:instr}.

Once we collected the ground truth, we ran our analysis scripts (privacy attacks) over the collected data to compile and infer data points, which we compared to the ground truth to assess the attacks' accuracy. The results of these experiments are described in \S\ref{sec:Experiment_results} and summarized in Table \ref{tab:results}.


\vspace{1em}

\subsection{Ethical considerations}
\label{subsec:ethical_considerations}

We identified three primary ethical risks in our protocol: (i) the risk of discomfort using a VR device, (ii) the risk of a confidentiality breach of participant data, and (iii) the risk that participants might not have wished to disclose certain information about themselves during the course of the study.

To address the first risk (i), we used high-fidelity VR devices and appropriately powerful gaming computers for all participants, together capable of consistently providing $120$ frames per second, well above the minimum specifications recommended to mitigate the risk of VR sickness \cite{nato_cybersickness}. We designed our VR game to avoid distressing elements such as horror, claustrophobia, or flickering/strobing lights. Furthermore, a researcher was present to ensure participants did not collide with real-world objects during each play-through.

\begin{figure*}[!ht]
    \centering
    \vspace{1em}
    \begin{subfigure}{.33\linewidth}
        \centering
        \includegraphics[width=\linewidth]{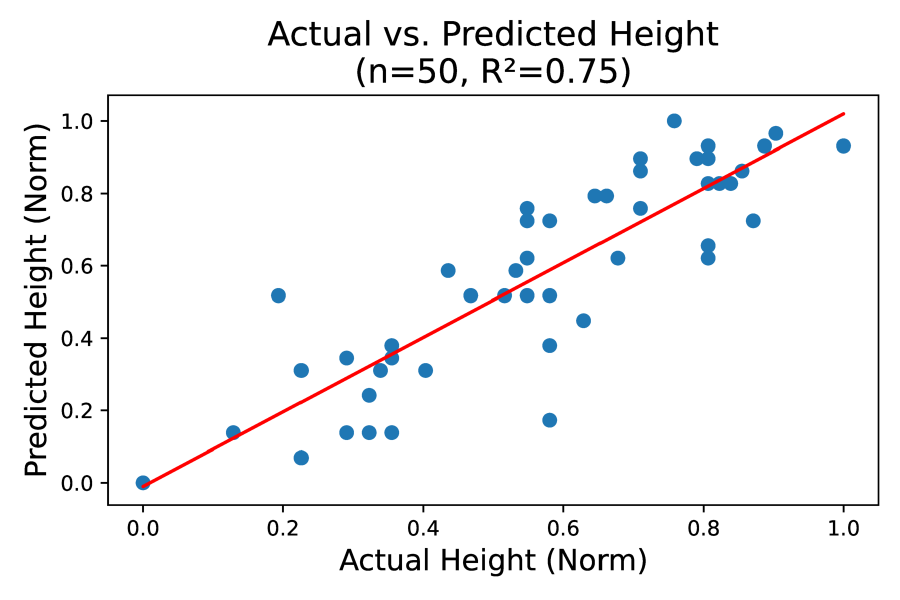}
        \label{fig:height}
    \end{subfigure}%
    \hfill
    \begin{subfigure}{.33\linewidth}
        \centering
        \includegraphics[width=\linewidth]{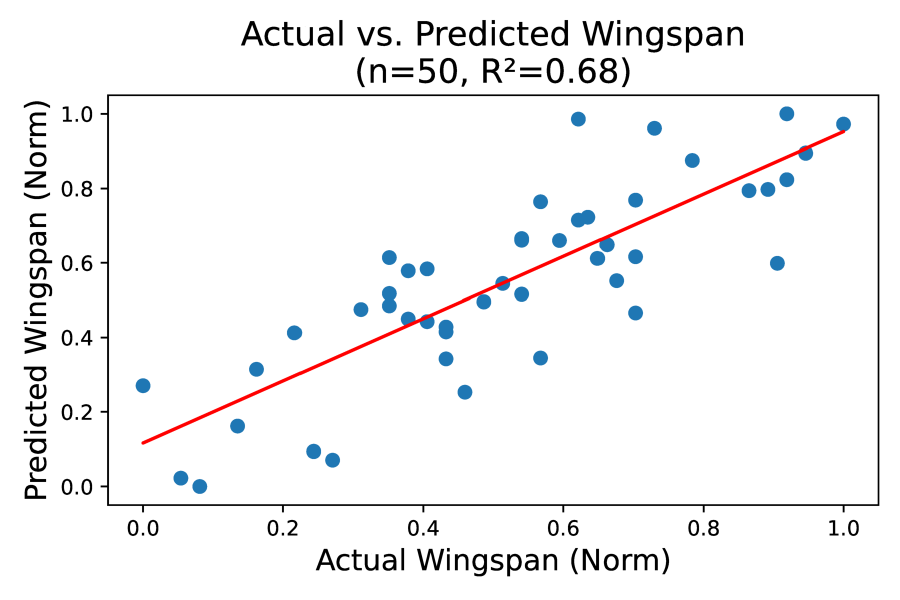}
        \label{fig:wingspan}
    \end{subfigure}%
    \hfill
    \begin{subfigure}{.33\linewidth}
        \centering
        \includegraphics[width=\linewidth]{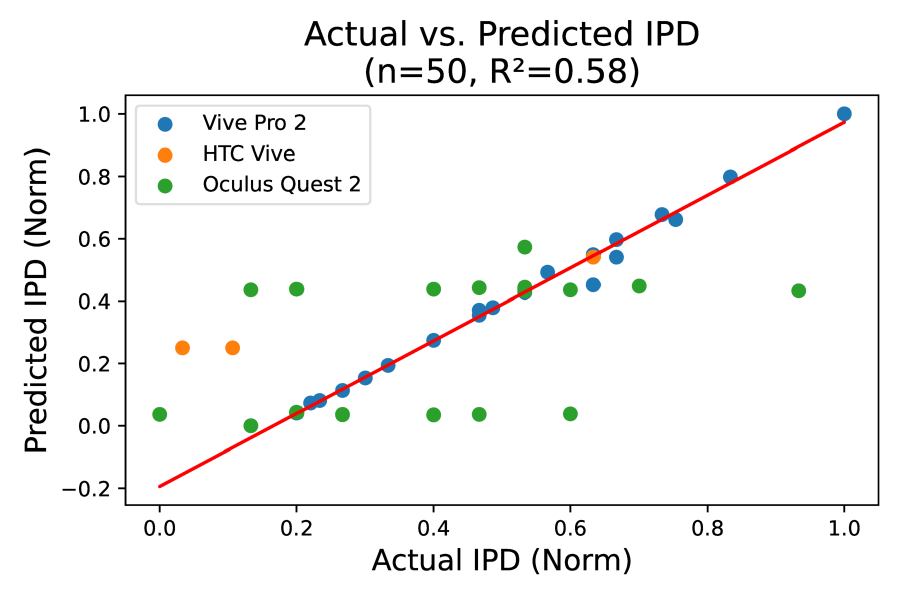}
        \label{fig:ipd}
    \end{subfigure}%
    \caption{Actual and predicted user anthropometrics.}
    \label{fig:anthropometrics_linear_Reg}
    \vspace{1em}
\end{figure*}

To address the second risk (ii), we anonymized all collected data using random codes that we could not reasonably trace back to a participant’s identity. Moreover, we avoided collecting any highly-sensitive data that could potentially damage participants in a breach. Lastly, we normalized biometric measurements on a scale of $0$ to $1$ to avoid revealing exact measurements in this paper (e.g., in Fig.~\ref{fig:anthropometrics_linear_Reg}).  The photos included in this paper are not of actual participants.

To address the third risk (iii), we made sure participants clearly understood the nature of the study. We emphasize that this is not a deception study. Our claims about the non-obviousness of the presented attacks should not be construed to imply that participants were unaware that their data was being collected during the study. Participants were informed that their data was being collected, including a description of the categories of data being observed. After completing the VR portion of the study, participants were made aware of the exact attributes being collected. They were explicitly given the opportunity to withdraw consent without penalty at any point in the process, including after having detailed knowledge of the data attributes involved, in which case their data would not have been included in the results.

In light of these considerations, the study was deemed a minimal-risk behavioral intervention and was granted an IRB exempt certification under 45 C.F.R. § 46.104(d)(3) by an OHRP-registered institutional review board.

\begin{table*}
\normalsize
\centering
\setlength{\tabcolsep}{6pt}
\resizebox{\textwidth}{!}{%
\begin{tabular}{|l|l|l|l|l|l|}
\hline
\textbf{Attribute} & \textbf{Type / Source} & \textbf{Precision} & \textbf{Accuracy} & \textbf{Statistics} & \textbf{Attackers} \\ \hline \hline

\textbf{Height} & \begin{tabular}[c]{@{}l@{}}Primary\\
Telemetry\end{tabular} & $1$ cm & \begin{tabular}[c]{@{}l@{}}76\% within $5$~cm\\ $94\%$ within $7$~cm\end{tabular} & $R^2=0.75$ & \begin{tabular}[c]{@{}l@{}}Privileged I-III\\ Non-Privileged*\end{tabular} \\ \hline

\textbf{Longer Arm} & \begin{tabular}[c]{@{}l@{}}Primary\\ Telemetry\end{tabular} & boolean & \begin{tabular}[c]{@{}l@{}}$58\%$ for $\geq2$~cm difference\\ $100\%$ for $\geq3$~cm difference\end{tabular} & \begin{tabular}[c]{@{}l@{}}$F_1=0.67$\\$F_1=1.00$\end{tabular} & \begin{tabular}[c]{@{}l@{}}Privileged I-III\\ Non-Privileged*\end{tabular} \\ \hline

\textbf{Interpupillary Distance} & \begin{tabular}[c]{@{}l@{}}Primary\\ Telemetry\end{tabular} & $0.1$~mm & \begin{tabular}[c]{@{}l@{}}$96\%$ within $0.5$~mm (Vive Pro 2)\\ $58\%$ within $0.5$~mm (All Devices)\end{tabular} & \begin{tabular}[c]{@{}l@{}}$R^2=0.99$\\$R^2=0.58$\end{tabular} & Privileged I-II \\ \hline

\textbf{Wingspan} & \begin{tabular}[c]{@{}l@{}}Secondary\\ Telemetry\end{tabular} & $1$~cm & \begin{tabular}[c]{@{}l@{}}78\% within $7$~cm\\ $98\%$ within $12$~cm\end{tabular} & $R^2=0.68$ & \begin{tabular}[c]{@{}l@{}}Privileged I-III\\ Non-Privileged*\end{tabular} \\ \hline

\textbf{Room Size} & \begin{tabular}[c]{@{}l@{}}Secondary\\ Telemetry\end{tabular} & $1$~m$^2$ &  \begin{tabular}[c]{@{}l@{}}70\% within $2$~m$^2$ \\ $96\%$ within $3$~m$^2$ \end{tabular} & $R^2=0.97$ & \begin{tabular}[c]{@{}l@{}}Privileged I-III\\ Non-Privileged*\end{tabular} \\ \hline

\textbf{Geolocation} & \begin{tabular}[c]{@{}l@{}}Primary\\ Network\end{tabular} & $100$~km & \begin{tabular}[c]{@{}l@{}}50\% within $400$~km\\ $94\%$ within $500$~km\end{tabular} & N/A & Privileged II-III \\ \hline

\textbf{HMD Refresh Rate} & \begin{tabular}[c]{@{}l@{}}Primary\\ Device\end{tabular} & $1$~Hz & \begin{tabular}[c]{@{}l@{}}$100\%$ within $3$~Hz (Privileged Attacker)\\ $88\%$ wtihin $60$~Hz (Unprivileged Attacker)\end{tabular} & \begin{tabular}[c]{@{}l@{}}$R^2=0.99$\\$R^2=0.75$\end{tabular} & \begin{tabular}[c]{@{}l@{}}Privileged I-II\\ Privileged III*\\ Non-Privileged*\end{tabular} \\ \hline

\textbf{Controller Tracking Rate }& \begin{tabular}[c]{@{}l@{}}Primary\\ Device\end{tabular} & $1$~Hz & $100\%$ within $2.5$~Hz & $R^2=0.99$ & \begin{tabular}[c]{@{}l@{}}Privileged I-II\\ Privileged III*\\ Non-Privileged*\end{tabular} \\ \hline

\textbf{Device Resolution (MP)} & \begin{tabular}[c]{@{}l@{}}Primary\\ Device\end{tabular} & $0.1$~MP & $100\%$ within $0.1$~MP & $R^2=1.00$ & Privileged I-II \\ \hline

\textbf{Device FOV} & \begin{tabular}[c]{@{}l@{}}Primary\\ Device\end{tabular} & $10$° & $100\%$ within $10$° & $R^2=0.92$ & \begin{tabular}[c]{@{}l@{}}Privileged I-II\\ Privileged III*\\ Non-Privileged*\end{tabular} \\ \hline

\textbf{Computational Power} & \begin{tabular}[c]{@{}l@{}}Primary\\ Device\end{tabular} & \begin{tabular}[c]{@{}l@{}}$0.1$~GHz\\ $10$~Mh/s\end{tabular} & \begin{tabular}[c]{@{}l@{}}CPU: 100\% within $0.4$~GHz\\ GPU: $100\%$ within $20$~Mh/s\end{tabular} & \begin{tabular}[c]{@{}l@{}}$R^2=0.92$\\$R^2=0.81$\end{tabular} & Privileged I-II \\ \hline

\textbf{VR Device} & \begin{tabular}[c]{@{}l@{}}Secondary\\ Device\end{tabular} & categorical & $100\%$ & $p=0.00$ & \begin{tabular}[c]{@{}l@{}}Privileged I-III\\ Non-Privileged*\end{tabular} \\ \hline

\textbf{Handedness} & \begin{tabular}[c]{@{}l@{}}Primary\\ Behavior\end{tabular} & boolean & $96\%$ & $F_1=0.98$ & \begin{tabular}[c]{@{}l@{}}Privileged I-III\\ Non-Privileged\end{tabular} \\ \hline

\textbf{Eyesight} & \begin{tabular}[c]{@{}l@{}}Primary\\ Behavior\end{tabular} & boolean & \begin{tabular}[c]{@{}l@{}}72\% (Hyperopia)\\ $80\%$ (Myopia)\end{tabular} & \begin{tabular}[c]{@{}l@{}}$F_1=0.73$\\$F_1=0.75$\end{tabular} & \begin{tabular}[c]{@{}l@{}}Privileged I-III\\ Non-Privileged\end{tabular} \\ \hline

\textbf{Color Blindness} & \begin{tabular}[c]{@{}l@{}}Primary\\ Behavior\end{tabular} & boolean & $100\%$ & $F_1=1.00$ & \begin{tabular}[c]{@{}l@{}}Privileged I-III\\ Non-Privileged\end{tabular} \\ \hline

\textbf{Languages} & \begin{tabular}[c]{@{}l@{}}Primary\\ Behavior\end{tabular} & boolean & $90\%$ & $p=0.08$ & \begin{tabular}[c]{@{}l@{}}Privileged I-III\\ Non-Privileged\end{tabular} \\ \hline

\textbf{Physical Fitness} & \begin{tabular}[c]{@{}l@{}}Primary\\ Behavior\end{tabular} & boolean & $86\%$ & $F_1=0.92$ & \begin{tabular}[c]{@{}l@{}}Privileged I-III\\ Non-Privileged\end{tabular} \\ \hline

\textbf{Reaction Time} & \begin{tabular}[c]{@{}l@{}}Primary\\ Behavior\end{tabular} & categorical & $88\%$ & $F_1=0.90$ & \begin{tabular}[c]{@{}l@{}}Privileged I-II\\ Privileged III*\\ Non-Privileged*\end{tabular} \\ \hline

\textbf{Acuity (MoCA)} & \begin{tabular}[c]{@{}l@{}}Primary\\ Behavior\end{tabular} & $1$~point & \begin{tabular}[c]{@{}l@{}}$94\%$ within $2$~points\\ $100\%$ diagnostic accuracy\end{tabular} & $F_1=1.00$ & \begin{tabular}[c]{@{}l@{}}Privileged I-III\\ Non-Privileged\end{tabular} \\ \hline

\textbf{Gender} & \begin{tabular}[c]{@{}l@{}}Inferred\\ Classification\end{tabular} & boolean & $98\%$ & $F_1=0.98$ & \begin{tabular}[c]{@{}l@{}}Privileged I-III\\ Non-Privileged\end{tabular} \\ \hline

\textbf{Age} & \begin{tabular}[c]{@{}l@{}}Inferred\\ Regression\end{tabular} & $1$~yr & $100\%$ within $1.5$~yr & $R^2=0.99$ & \begin{tabular}[c]{@{}l@{}}Privileged I-III\\ Non-Privileged\end{tabular} \\ \hline

\textbf{Ethnicity} & \begin{tabular}[c]{@{}l@{}}Inferred\\ Classification\end{tabular} & categorical & $98\%$ & $p=0.01$ & \begin{tabular}[c]{@{}l@{}}Privileged I-III\\ Non-Privileged\end{tabular} \\ \hline


\textbf{Disability Status} & \begin{tabular}[c]{@{}l@{}}Inferred\\ Classification\end{tabular} & boolean & $100\%$ & $F_1=1.00$ & \begin{tabular}[c]{@{}l@{}}Privileged I-III\\ Non-Privileged\end{tabular} \\ \hline

\textbf{Identity} & \begin{tabular}[c]{@{}l@{}}Inferred\\ Classification\end{tabular} & categorical & $100\%$ & $p=0.00$ & \begin{tabular}[c]{@{}l@{}}Privileged I-III\\ Non-Privileged\end{tabular} \\ \hline
\end{tabular}} \\
\vspace{1em}
\centering{\normalsize * With degraded accuracy.}

\caption{Selected attributes collected and analyzed during the experiment, with accuracy and $R^2$, $F_1$, or $p$ values from $\chi^2$ tests.}
\label{tab:results}
\end{table*}

\section{Results}
\label{sec:Experiment_results}

In this section, we present the empirical effectiveness of the privacy attacks introduced in \S\ref{sec:Privacy_attacks}, as summarized in Table~\ref{tab:results}.

\subsection{Biometrics}
\label{sec:Biometrics}

\noindent{\textbf{Continuous Anthropometrics}.}
Fig.~\ref{fig:anthropometrics_linear_Reg} shows (scaled) actual and predicted values for \textit{height} ($R^2=0.75$), \textit{wingspan} ($R^2=0.68$), and \textit{interpupillary distance} (IPD) ($R^2=0.58$). 
IPD measurements were most accurate on the Vive Pro 2, with $R^2=0.99$ when excluding other devices. 
In general, we could accurately determine these three metrics for most users from just a few seconds of telemetry. 
We were not, however, able to accurately predict the individual lengths of the left and right arms ($R^2=0.02$ and $R^2=0.01$ respectively), due to the lack of a reliable center point from which to measure.\smallskip

\noindent{\textbf{Binary Anthropometrics}.}
Although absolute arm lengths were not discernible, relative lengths were accurate enough that we could usually identify which of the participant's arms was longer. We observed increasing accuracy for participants with greater differences in length, reaching $100\%$ accuracy for the $12\%$ of participants with a difference of at least $3$~cm. Handedness can also be determined accurately from behavioral observations; we note, however, that $94\%$ of our participants reported being right-handed.\smallskip

\noindent{\textbf{Vision}.}
Our vision tests achieved diagnostic accuracies for \textit{hyperopia} (farsightedness), \textit{myopia} (nearsightedness), and \textit{deuteranopia} (red-green color blindness) of $72\%$, $80\%$, and $100\%$ respectively. The overall accuracy of detecting a visual deficiency was $80\%$, in part because some users of contact lenses could not remove their contacts for the experiment. \smallskip

\noindent{\textbf{Fitness}.}
Using squat depth as a correlate of \textit{physical fitness} discriminated ``low" fitness with an accuracy of $86\%$; our tests were not able to differentiate between ``moderate" and ``high" fitness.\smallskip

\noindent{\textbf{Reaction Time}.}
We measured \textit{reaction time} to a precision of one recorded frame ($16.6$~ms). We were able to detect whether a participant's reaction time was above or below $250$~ms (the approximate median reaction time) with an accuracy of $88\%$.

\subsection{Environment}
\noindent{\textbf{Room Size}.} The \textit{length} and \textit{width} of each of three testing rooms was determined to within $1.0$~m with accuracies of nearly $90\%$. This allowed true room area to be found within $3$~m$^2$ in $96\%$ of trials. Taking the average estimated area for each tested room vs. the true accessible room area yields $R^2=0.97$.\smallskip

\noindent{\textbf{Geolocation}.} Using the server latency multilateration (hyperbolic positioning) technique for \textit{geolocation} yielded a mean longitudinal and latitudinal error of around 2.5° across four tested locations. This was sufficient to locate the test subject to within $500$~km in $94\%$ of cases, and within the correct state in $100\%$ of cases.

\subsection{Device Specifications}

\noindent{\textbf{Tracking Rate}.} We found that privileged attackers could determine various VR device specifications (namely, \textit{display refresh rate}, \textit{display resolution}, \textit{field of view}, and \textit{tracking rate}) with high accuracy. Tracking rate (the number of unique telemetry measurements taken per second) is a particularly interesting metric, as the top four VR headsets, together accounting for over 75\% market share \cite{noauthor_valve_2021}, all have different default HMD refresh rates (72/144/80/90 Hz).\smallskip

\noindent{\textbf{VR Device}.} Using the above device specifications and the highly heterogeneous nature of VR device specifications, privileged attackers can determine the type of VR device with $100\%$ accuracy. We also found that non-privileged attackers could determine the refresh rate to within $30$~Hz with an accuracy of $30\%$ and to within $60$~Hz with an accuracy of $88\%$; however, this was not sufficient to accurately determine the type of device.\smallskip

\noindent{\textbf{Host Device}.} We found that an attacker benchmarking host device specifications can determine \textit{GPU power} with $100\%$ accuracy to within $20$~Mh/s (daggerhashimoto) and \textit{CPU clock speed} to within $0.4$~GHz, allowing them to estimate the price tier of the host device.

\subsection{Acuity (MoCA)}
\label{sec:Acuity}
Table~\ref{tab:moca} summarizes the numerical (continuous, i.e., the score of each category) and diagnostic (binary, i.e., passing or failing a category) accuracy of the \textit{Montreal Cognitive Assessment} (MoCA) we conducted in the VR experiments. We achieved a diagnostic accuracy of $90\%$ or greater for $5$ of the $7$ scored MoCA categories (excluding visuospatial/executive and delayed recall), with an overall diagnostic accuracy of $100\%$.

\begin{table}
\setlength{\tabcolsep}{8pt}
\centering
\resizebox{0.75 \columnwidth}{!}{%
\begin{tabular}{|l|l|l|}
\hline
\textbf{\begin{tabular}[c]{@{}l@{}}MoCA\\ Category\end{tabular}} & \textbf{\begin{tabular}[c]{@{}l@{}}Accuracy\\ (Numerical)\end{tabular}} & \textbf{\begin{tabular}[c]{@{}l@{}}Accuracy\\ (Diagnostic)\end{tabular}} \\ \hline\hline
\textbf{Executive} & N/A & N/A \\ \hline
\textbf{Naming} & $100$\% & $100$\% \\ \hline
\textbf{Memory} & $78$\% & $84$\% \\ \hline
\textbf{Serial $7$} & $90$\% & $100$\% \\ \hline
\textbf{Attention} & $88$\% & $100$\% \\ \hline
\textbf{Repetition} & $74$\% & $96$\% \\ \hline
\textbf{Language} & $74$\% & $96$\% \\ \hline
\textbf{Abstraction} & $100$\% & $100$\% \\ \hline
\textbf{Recall} & $60$\% & $90$\% \\ \hline
\textbf{Orientation} & $100$\% & $100$\% \\ \hline\hline
\textbf{Overall} & \begin{tabular}[c]{@{}l@{}}80\% within $1$ point\\ $94$\% within $2$ points\end{tabular} & $100$\% \\ \hline
\end{tabular}%
}
\caption{Accuracy of each MoCA category.}
\label{tab:moca}
\end{table}

\subsection{Demographics}
\label{sec:Demographics}
\noindent{\textbf{Language}.} The visual focus method of \textit{language} determination identified a spoken language (other than English) with at least conversational proficiency in $90\%$ of multilingual participants. \smallskip

\noindent{\textbf{Vocal Characteristics}.} We used existing machine learning models to determine the gender~\cite{becker_primaryobjectsvoice-gender_2022} and ethnicity~\cite{garg_speech-accent-recognition_2022} of participants from their voice with an accuracy of $98\%$ and $66\%$ respectively; these accuracy values improved to $100\%$ when combined with other attributes such as height and wingspan as described in ``Inferred Attributes'' below.\smallskip

\noindent{\textbf{Inferred Attributes}.} We used Azure Automated Machine Learning~\cite{azure_ml} to determine the optimal preprocessor, model type, and input metrics for inferring several demographic attributes. Table~\ref{tab:inference} summarizes the results of this meta-analysis. For identity, we used the best-performing technique of Miller et al. \cite{miller_personal_2020}. Using the identified optimal models, we determined the participant's gender, ethnicity, disability status, age (within 1.5 years), and identity with nearly $100\%$ accuracy across several Monte Carlo cross-validations; thus, users were never simultaneously present in the training and testing datasets, other than for inferring identity, and it is not possible that the demographic inferences were a result of identifying users.

With respect to disability, we there was one reported physical disability and three reported mental disabilities amongst our $50$ participants; we were able to identify these disabilities individually with $100\%$ accuracy ($F_1=1.00$). In each case, the model far outperformed any individual attribute; for example, ethnicity was $98\%$ accurate despite its most significant input (voice) being only $66\%$ accurate on its own.

\begin{table}[H]
\resizebox{\columnwidth}{!}{%
\begin{tabular}{|l|l|l|}
\hline
\textbf{Attribute (Prediction)} & \textbf{Inputs} & \textbf{Preprocessing / Model} \\ \hline \hline

\begin{tabular}[c]{@{}l@{}}\textbf{Gender}\\ (Classification)\end{tabular} & \begin{tabular}[c]{@{}l@{}}Voice, Height, Wingspan,\\ Interpupillary Distance (IPD)\end{tabular} & \begin{tabular}[c]{@{}l@{}}TruncatedSVDWrapper\\ SVM\end{tabular} \\ \hline

\begin{tabular}[c]{@{}l@{}}\textbf{Age}\\ (Regression)\end{tabular} & \begin{tabular}[c]{@{}l@{}}Close Vision, Reaction Time,\\ Height, Test Duration, Acuity\end{tabular} & \begin{tabular}[c]{@{}l@{}}MaxAbsScaler\\ ExtremeRandomTrees\end{tabular} \\ \hline

\begin{tabular}[c]{@{}l@{}}\textbf{Ethnicity}\\ (Classification)\end{tabular} & Voice, Language, Height & \begin{tabular}[c]{@{}l@{}}StandardScalerWrapper\\ LightGBM\end{tabular} \\ \hline 


\begin{tabular}[c]{@{}l@{}}\textbf{Disabilities}\\ (Classification)\end{tabular} & Vision, Fitness, Acuity & \begin{tabular}[c]{@{}l@{}}MaxAbsScaler\\ NaiveBayes\end{tabular} \\ \hline

\begin{tabular}[c]{@{}l@{}}\textbf{Identity}\\ (Classification)\end{tabular} &  \begin{tabular}[c]{@{}l@{}}Height, Wingspan, Acuity,\\IPD, Vision, Reaction Time\end{tabular} & \begin{tabular}[c]{@{}l@{}}StandardScalerWrapper\\ RandomForest\end{tabular} \\ \hline

\end{tabular}%
}
\caption{Inputs and methodology of inferred attributes.}
\label{tab:inference}
\end{table}

\eject

By following the Azure Automated ML Workflow~\cite{azure_ml}, we avoided the biases of manually selecting features for demographic inference.

The AutoML workflow begins by clustering users into training, validation, and testing sets. Using the validation set, a variety of preprocessing techniques are first evaluated. Next, a variety of classical machine learning models are trained using the preprocessed features, using the validation set to evaluate the accuracy of each algorithm. For the most accurate architecture, a hyperparameter sweep is performed to optimize the model for the feature set.

Using the best-performing model, an explainability analysis is conducted to select the most important input features for inferring each attribute.
For example, rather than instructing the model to infer gender from voice, height, wingspan, etc., we initially provided the system with all available primary attributes, and allowed it to determine on its own (e.g., via PCA) which are relevant to a given inference task. Finally, the testing set is used to evaluate the selected approach, consisting of automatically-determined features, preprocessing, architecture, and hyperparameters. The process is repeated across several Monte Carlo cross-validations
\section{Discussion}
\label{sec:Discussion}

We now return to the question of whether an attacker can use data collected from consumer-grade VR devices to extract and infer users' private information accurately.
In this study, we have shown that this is indeed possible, with moderate to high accuracy values for most of the aggregated and inferred data points presented in Table~\ref{tab:results}.
Using the technique of Miller et al. \cite{miller_personal_2020}, we found that an attacker could consistently identify a participant among the pool of $50$ within a few minutes of gameplay.
Moreover, we have collected more than $25$ granular data points, well above the $15$ necessary to uniquely identify every individual in the United States \cite{rocher}.
While we were required to condense this data collection into a concise $20$-minute experiment for logistical reasons, real-world attackers could gain increased accuracy and covertness by aggregating data collected over much longer periods of time.

\subsection{Participant Awareness}
In sections~\ref{sec:Privacy_attacks} and~\ref{sec:Experimental_Design}, we argued that a developer could design VR environments and games to facilitate the covert collection of targeted data points disguised as normal game elements.

For ethical reasons, our participants were informed that they were participating in an adversarial experiment where their personal information would be collected. However, they were not told exactly which attributes were being collected until the end of the experiment. Thus, our experiment reflects a realistic scenario in which users may be generally aware of privacy threats but would not necessarily know the attributes collected by a given application.

Upon debriefing, we asked the participants whether they could identify which of the attributes we were attempting to measure. All $50$ participants reported not knowing exactly which attributes were being collected and inferred during the game, with no participant able to correctly identify more than $3$ of the attributes.

When we revealed the list of attributes, many participants expressed surprise at the breadth of information that could be collected within VR, but none expressed particular shock at the existence of some degree of data harvesting (perhaps having already grown accustomed to these practices on the web). Even if participants were aware of the attributes being measured, we believe it would have been difficult to counteract many of the attacks, due to the deeply subconscious nature of many of the observed behaviors.

\subsection{Adversarial Capabilities}
In \S\ref{sec:biom}, we provided wingspan as an initial example of an attribute that is far easier to observe in an adversarial application; while users rarely adopt a posture in which their arms are completely outstretched, an adversarial game can easily drive users to adopt such a posture through the use of an interactive element.

Reflecting now on the entire set of attributes inferred in this study, we observe that nearly all of them are aided by the introduction of adversarial puzzles in our escape room. For example, most behavioral observations, including the entire MoCA acuity assessment, relied on observing the user's responses to specific adversarially-constructed puzzles.

Our major conclusion from this finding is that the privacy risk of VR devices stems not only from their sensors, such as accelerometers and gyroscopes, but also from the immersive nature of their displays, which can be used to totally control a user's virtual environment to influence the information they reveal.
Thus, while past research has primarily focused on the capability to infer various attributes from motion-tracking data, it is equally important to consider the threat posed by adversarial VR application design.

\subsection{Societal Implications}
While for ethical reasons we limited our attacks to relatively benign data points, an attacker could potentially track and infer additional information about other more critically sensitive personality traits, like sexual, religious, or political orientation, educational level, and illnesses, among others, to enhance practices such as surveillance advertisement~\cite{surveillance_ads} or pushing political agendas~\cite{political_targeting}.
Given how immersive and emotionally engaging VR environments
can be~\cite{nelson_virtual_nodate, kerr_criminal_nodate, warren_crime_nodate, VR_priv_tutorial}, such practices could become more pernicious and effective than with current mobile and desktop applications.

The risk of profiling users in VR is further exacerbated by the known ability to uniquely identify users based on their VR motion data \cite{nair2023unique}. This already presents a significant privacy risk, as typical VR usage today includes rowdy gaming sessions and adult experiences that users may not want to be linkable to their identity in more professional settings. Further, by combining our profiling techniques with cross-application identification methods, an attacker, or even a group of colluding attackers, could attempt to aggregate user profiles from data across many applications.

Although we use the terms ``attack" and ``attacker" throughout this paper, to the best of our knowledge, there is nothing illegal about the methods described herein. It is possible that in the future, many VR users would knowingly or unknowingly consent to this form of data collection via clauses contained in platform terms of service or end-user license agreements.
In fact, major VR device manufacturers have been observed selling headsets at a loss~\cite{published_oculus_2021} of up to \$$10$~billion per year~\cite{published_despite_2022}, evidently with the aim to recoup said losses with some form of after-sales revenue.

\eject

\subsection{Limitations}
The results of this study should be understood in light of a few limitations.
Unfortunately, our sample of participants was not perfectly representative of the general population; for example, college students were overrepresented.
For logistical reasons, we were unable to tamper with VR device firmware and thus could not consider hardware-level attacks in this paper. Therefore, privileged attackers I and II, while different in theory, had identical capabilities within the scope of our experiment. While ground truth values for user attributes were measured by the researchers when possible, many were also self-reported (see \S\ref{app:instr}) and thus potentially biased.
Lastly, the researchers were forced to interact with participants outside of VR on some occasions, such as to warn of nearby obstacles. While we did attempt to minimize such occurrences, these interactions could nevertheless have biased certain results. \medskip

\subsection{Future Work}
Given the early stage of research on privacy and VR, there are many outstanding questions in this field for researchers to tackle.
Among them is the question of how developers can design VR games or applications that make privacy attacks even more stealthy, including by integrating these attacks into daily tasks in future VR/AR environments.
On the other hand, researchers could also study analysis techniques for revealing hidden data collection mechanisms (where possible) to make these attacks harder to achieve.
Furthermore, studying what additional data attributes an attacker could leverage from data sources we did not consider (including eye tracking and full-body tracking) will expand our overall awareness of VR-related vulnerabilities.
Additionally, future work dedicated to how an attacker could not just observe but actually change users' opinions will shed light on the implications of future immersive and impactful metaverse applications.

Above all, we believe it is most important to study the potential countermeasures to these VR privacy attacks. Unfortunately, it is difficult to defend against attacks based on inferring private attributes from VR motion data. In VR, motion data is often required for legitimate purposes, such as passive authentication or multiplayer functionality. Thus, it is difficult for a user to ascertain whether their data is being used for malicious or benign purposes without auditing the server-side logic. Nevertheless, we can propose a few broad approaches for combatting this threat. \medskip

\noindent \textbf{Local Differential Privacy.} In ``MetaGuard'' \cite{nair2022going}, researchers have demonstrated that differentially private noise sampling functions, such as the bounded Laplace mechanism, are reasonably effective at defeating VR privacy attacks. While this approach incurs a usability penalty, it does so according to a theoretically optimal trade-off between data privacy and perceived tracking error. \medskip

\noindent \textbf{Adversarial Models.} Outside of VR, researchers have used generative adversarial networks (GAN) to obscure sensitive data from biometric sources like gait \cite{gan}.
While this approach loses the provable properties of differential privacy, thes models could actually be more effective at protecting privacy in practice, due to their ability to detect and obscure not only primary sensitive attributes but also hidden correlations to these variables. For example, if instructed to hide a user's height, the model would quickly learn to also obscure their wingspan, which is highly correlated to the former. \medskip

\noindent \textbf{Behavioral Modification.} ``Homuncular flexibility'' refers to the innate ability of users to operate control schemes that are very different from their own bodies. For example, VR users can effectively control avatars with 10 times their normal arm length, avatars with 3 arms, or even avatars representing completely different creatures, such as crabs \cite{10.1111/jcc4.12107}. In doing so, they may be less likely to introduce subconscious movement patterns that reveal personal information, than when controlling an anatomically familiar avatar.
Thus, homuncular flexibility could be leveraged to induce users to behave in ways that limit identifiability and profiling. \medskip

\noindent \textbf{Trusted Execution Environments.} Finally, there will likely remain some scenarios in which no amount of modification to tracking data is permissible. For example, in a competitive VR e-sports title, any perturbation of telemetry data could upset the scoring mechanism of the game. In such situations, trusted execution environments, such as Intel's SGX, may provide a hardware-based attestation mechanism that allows users to verify the software running on a remote machine before sending their data to that server. Thus, without modifying their motion data, users can ensure that only legitimate and necessary operations are being performed.


\section{Conclusion}
\label{sec:Conclusion}

In this study, we shed light on the serious privacy risks of the metaverse by showing how VR can be turned against its users.
Specifically, we provided a comprehensive security and privacy framework for VR environments that classifies (i) attackers, (ii) data sources, (iii) vulnerable attributes, and their corresponding (iv) attacks.
We demonstrated the practicality and accuracy of these attacks by designing and conducting experiments with $50$ participants using consumer-grade VR devices.
The participants played our ``escape room'' VR game, which was secretly designed to collect personal information, like biometrics, demographics, and VR device and network details, among numerous other data points.
The results demonstrate high information leakage with moderate to high accuracy values over most identified vulnerable attributes, with just a handful of these attributes being sufficient to uniquely identify a user~\cite{rocher_estimating_2019, sweeney_simple_nodate, narayanan_robust_2008, gao_elastic_2014, kondor_towards_2020, dwork_exposed_2017, sweeney_identifying_2013, archie_anonymization_nodate}.

The alarming accuracy and covertness of these attacks and the push of data-hungry companies towards metaverse technologies indicate that data collection and inference practices in VR environments will soon become more pervasive in our daily lives. 
Furthermore, the breadth of possible VR applications, increasing quality of VR devices, and relative simplicity of our demonstration, all suggest that more sophisticated attacks with a higher success rate are possible and perhaps on the horizon.
Therefore, we hope our work encourages other privacy practitioners to advance research at the intersection of privacy and VR, in particular to propose countermeasures for new and existing privacy attacks in the metaverse.

\eject

\section*{Availability}
The Unity (C\#) source code and compiled binaries for the ``escape room" VR game we designed for our experiments are available for download from our anonymized repository. The repository also contains the data collection instruments and data analysis scripts used to determine the primary attributes and sample data. \\

\noindent \centerline{\url{https://github.com/MetaDataStudyAnonymized/MetaDataStudy}}

\begin{acks}
This work was supported in part by the National Science Foundation, by the National Physical Science Consortium, and by the Fannie and John Hertz Foundation.
Any opinions, findings, and conclusions or recommendations expressed in this material are those of the authors and do not necessarily reflect the views of the supporting entities.
We sincerely thank our lab assistant, Ines Bouissou, and our study participants for making this work possible.
\end{acks}

\bibliographystyle{plain}
\bibliography{999_REFS}

\appendix
\section{Experimental Design}
\label{sec:VR_puzzles}

This section describes the experiment design in detail.
Our experiment consists of puzzles located in VR rooms that the participants visit.
The puzzles are artifacts that facilitate collecting privacy-sensitive variables that might not otherwise be evident.
The rooms are themed as a virtual office. Before initiating the game, we explained to the participants that they would find the password by solving a puzzle, thereby ``escaping'' the room.
As developing a full-fledged game with voice recognition or virtual password pads is out of scope, the participants spoke the passwords aloud so that the researchers could press a key and ``teleport'' them to the next room.
We include five ``noisy" rooms, i.e., rooms that do not serve the purpose of facilitating the measurement of sensitive information but help to mask the rooms that do.
Nonetheless, noisy rooms habituate the player to the game mechanics, e.g., looking around the room or immersing the player further in the game.
If the player gets stuck in one room, we press a key to teleport the participant to the next room.
We request the users to remove their glasses or contact lenses for puzzles $23$ and $24$, measuring eyesight.
While influencing players in such a way is not possible in a real scenario, these puzzles could at least identify the players who do not have good eyesight, i.e., they do not wear glasses/contacts when playing.

\graphicspath{{1100_Rooms/}}
\setlength{\abovecaptionskip}{0.5em}
\setlength{\belowcaptionskip}{0em}
\setlength{\floatsep}{0em}

\begin{figure}[H]
    \centering
    \includegraphics[width=\linewidth]{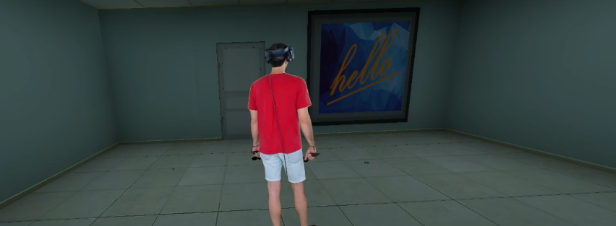}
    \caption*{\textbf{Puzzle 1}: The first room introduces the player to the dynamics of the game, containing only a door and a poster with the word ``\textit{hello}'', which is the password. Upon instinctively reading the word aloud, the player is teleported to the next room.}
\end{figure}

\begin{figure}[H]
    \centering
    \includegraphics[width=\linewidth]{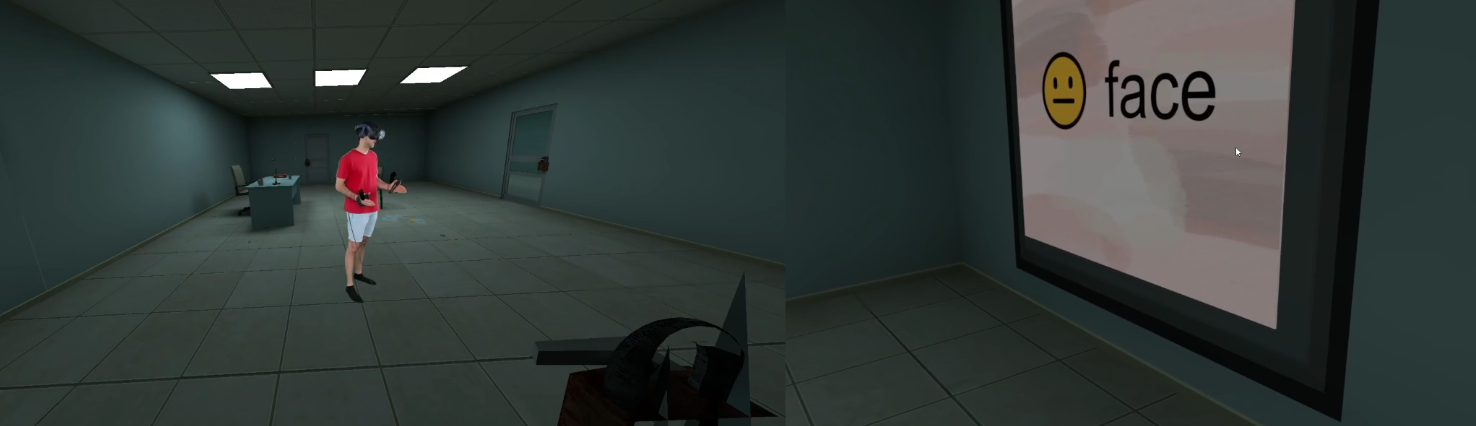}
    \caption*{\textbf{Puzzle 2}: The second room contains a poster with the password ``\textit{face}''. The player spawns facing the opposite wall of the poster; thus, we accustom the player to turn and explore the virtual environment to find the password and reinforce finding and speaking the password aloud.}
\end{figure}

\begin{figure}[H]
    \centering
    \includegraphics[width=\linewidth]{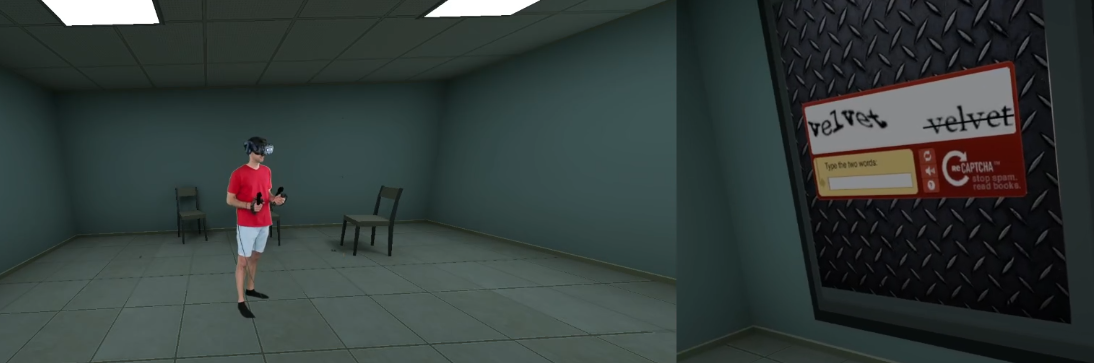}
    \caption*{\textbf{Puzzle 3}: Similarly, a poster depicts a captcha with the word ``\textit{velvet}.''}
\end{figure}

\begin{figure}[H]
    \centering
    \includegraphics[width=\linewidth]{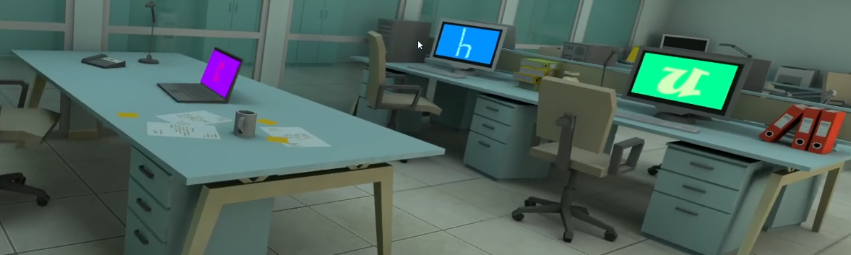}
    \caption*{\textbf{Puzzle 4}: The room contains several tables with monitors, on whose screens are letters spelling ``\textit{church}'' appropriately ordered from left to right.}
\end{figure}

\begin{figure}[H]
    \centering
    \includegraphics[width=\linewidth]{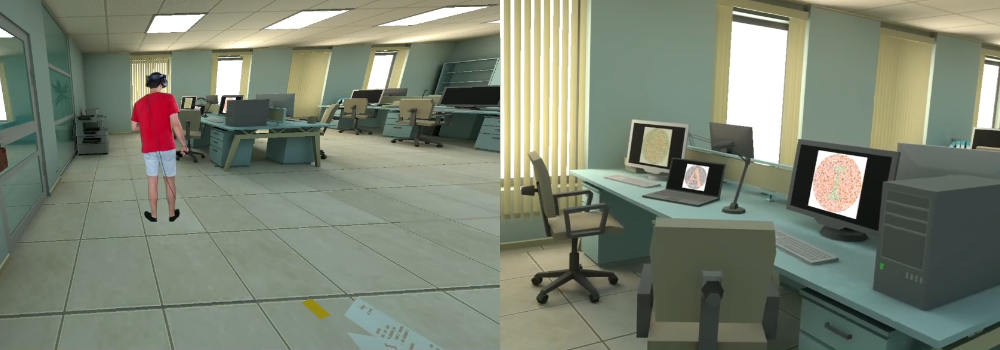}
    \caption*{\textbf{Puzzle 5}: This room tests for color blindness. Similarly to puzzle $4$, monitors display letters on Ishihara color test plates. Without colorblindness, the player would read ``\textit{daisy}''; with colorblindness, the player would read ``\textit{as}'' instead. Each of these passwords unlocks the room.}
\end{figure}

\begin{figure}[H]
    \centering
    \includegraphics[width=\linewidth]{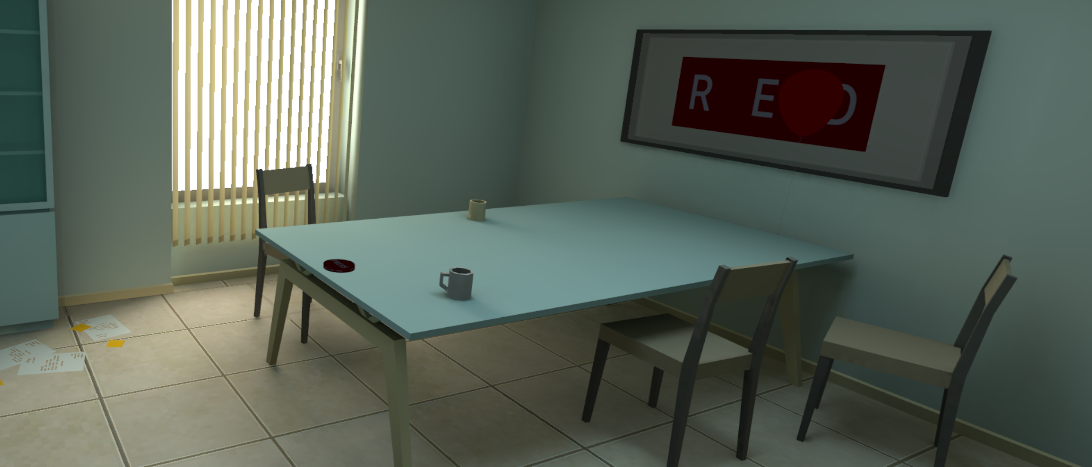}
    \caption*{\textbf{Puzzle 6}: There is a button on a table; upon pressing it three times, the three balloons next to the opposite wall pop sequentially, revealing the password ``\textit{red}''.}
\end{figure}

\begin{figure}[H]
    \centering
    \includegraphics[width=\linewidth]{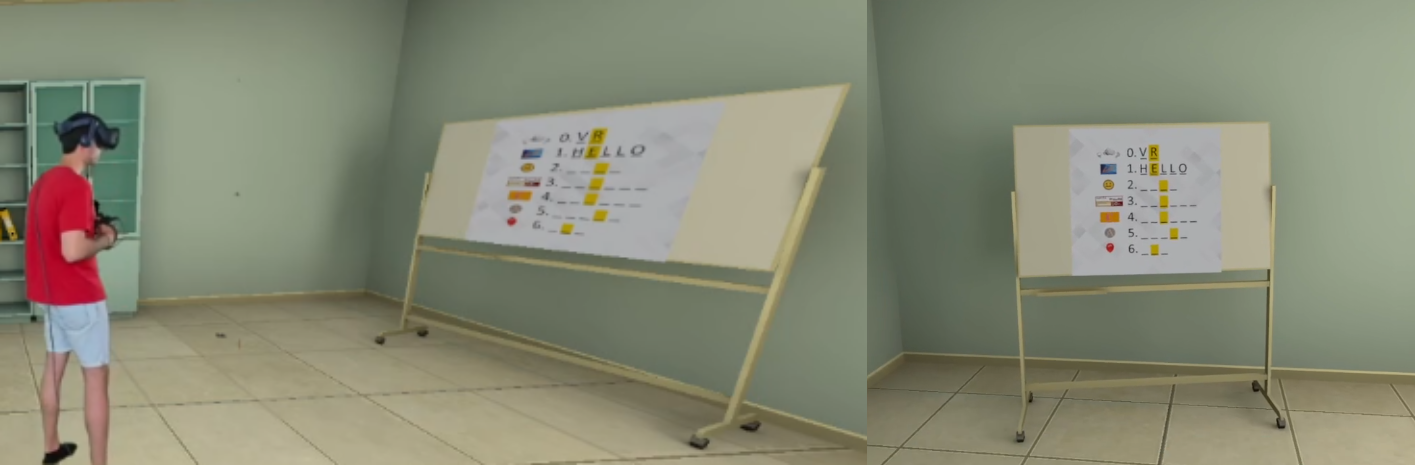}
    \caption*{\textbf{Puzzle 7}: The puzzle tests the short-term memory of the participants (MoCA memory). A whiteboard displays seven rows arranged vertically, each with fill-in blanks. The first two rows contain the already filled-in words ``\textit{VR}'' and ``\textit{hello}'', respectively. The last five rows correspond to the previous passwords from puzzles $2$ to $6$. Connecting the highlighted letters sequentially from up to bottom, the participant reveals the password ``\textit{recluse}''.}
\end{figure}

\begin{figure}[H]
    \centering
    \includegraphics[width=\linewidth]{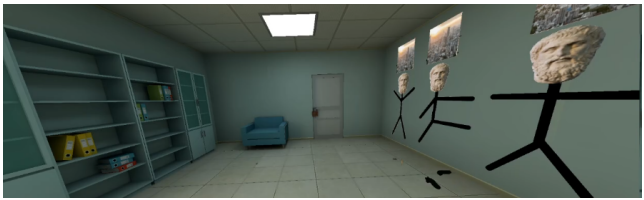}
    \caption*{\textbf{Puzzle 8}: To measure wingspan, we depict on a wall four human stick figures with different poses. The participant must mimic the poses on the wall to uncover the four letters of the password ``\textit{cave}''. One of the poses is a T-pose, which facilitates wingspan measurement.}
\end{figure}

\begin{figure}[H]
    \centering
    \includegraphics[width=\linewidth]{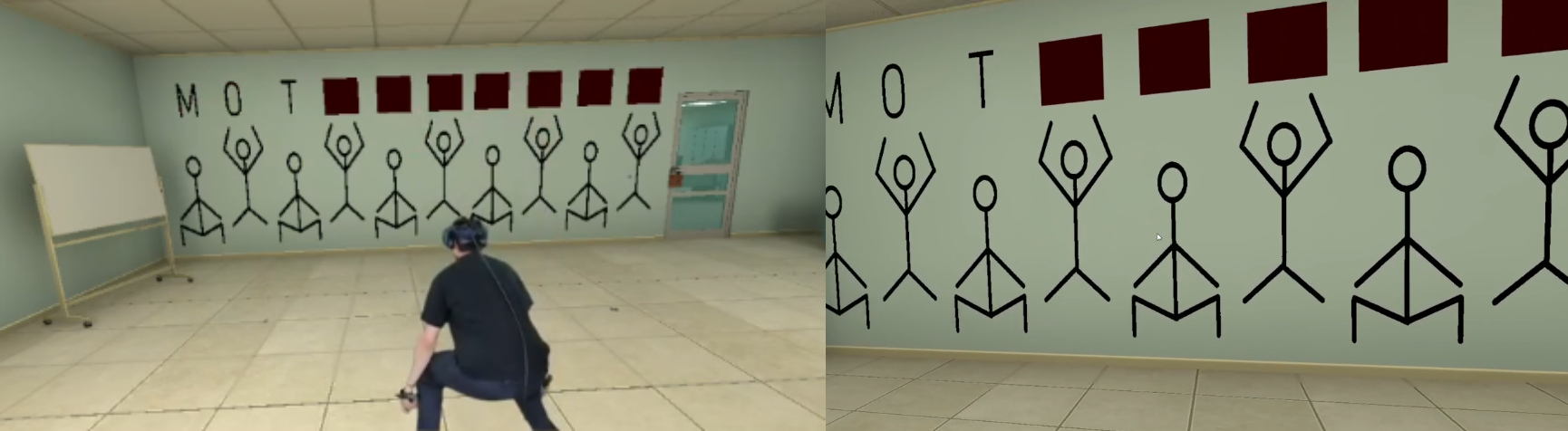}
    \caption*{\textbf{Puzzle 9}: The participant must mimic the sequence of poses on the wall, a set of squats. For every squat, the participant uncovers two letters of the password ``\textit{motivation}''. We correlate the distance traveled during the squats to fitness.}
\end{figure}

\begin{figure}[H]
    \centering
    \includegraphics[width=\linewidth]{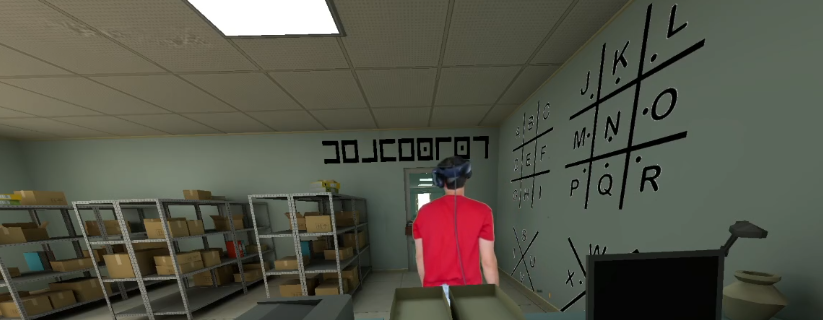}
    \caption*{\textbf{Puzzle 10}: The (noisy) room depicts on a wall a pigpen cipher hiding the password ``\textit{deafening}''.}
\end{figure}

\begin{figure}[H]
    \centering
    \includegraphics[width=\linewidth]{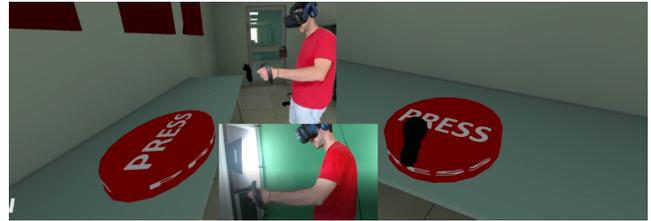}
    \caption*{\textbf{Puzzle 11}: The player presses a button on a table in time with a visual input, thereby revealing their reaction time.}
\end{figure}

\begin{figure}[H]
    \centering
    \includegraphics[width=\linewidth]{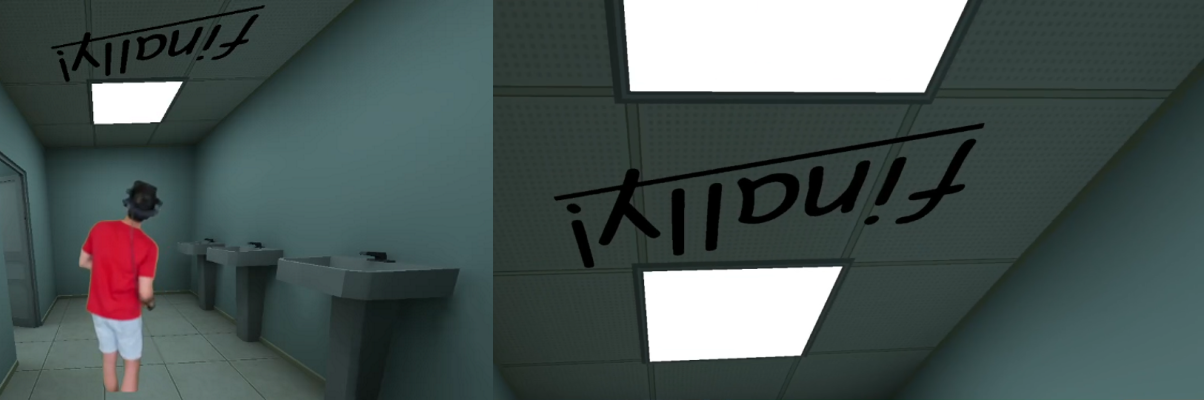}
    \caption*{\textbf{Puzzle 12}: The (noisy) room presents the password ``\textit{finally}'' on the ceiling, habituating the user to look also upwards.}
\end{figure}

\begin{figure}[H]
    \centering
    \includegraphics[width=\linewidth]{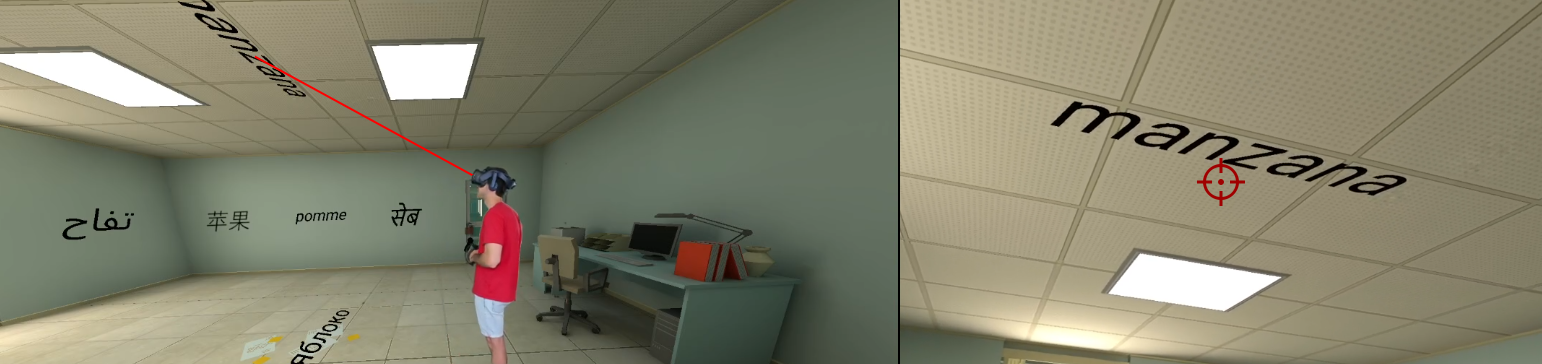}
    \caption*{\textbf{Puzzle 13}: The room depicts the word ``\textit{apple}'' in Hindi, Mandarin, French, Japanese, Russian, Spanish, Portuguese, and Arabic. The direction of gaze of the player when speaking the password reveals which language the participant recognizes.
}
\end{figure}

\begin{figure}[H]
    \centering
    \includegraphics[width=\linewidth]{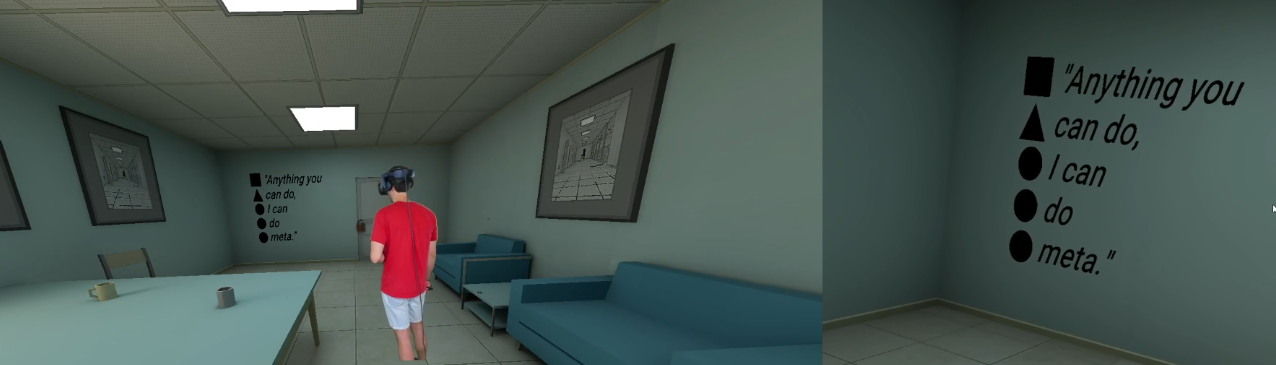}
    \caption*{\textbf{Puzzle 14}: This (noisy) room presents the sentence ``\textit{Everything you can do, I can do meta}'' broken down vertically into five rows. To the left of each row, there is a shape. The last three shapes are the same (circles). To solve the puzzle, the participant must read aloud the words next to the first instance of the repeated shape ``\textit{I can}.''}
\end{figure}

\begin{figure}[H]
    \centering
    \includegraphics[width=\linewidth]{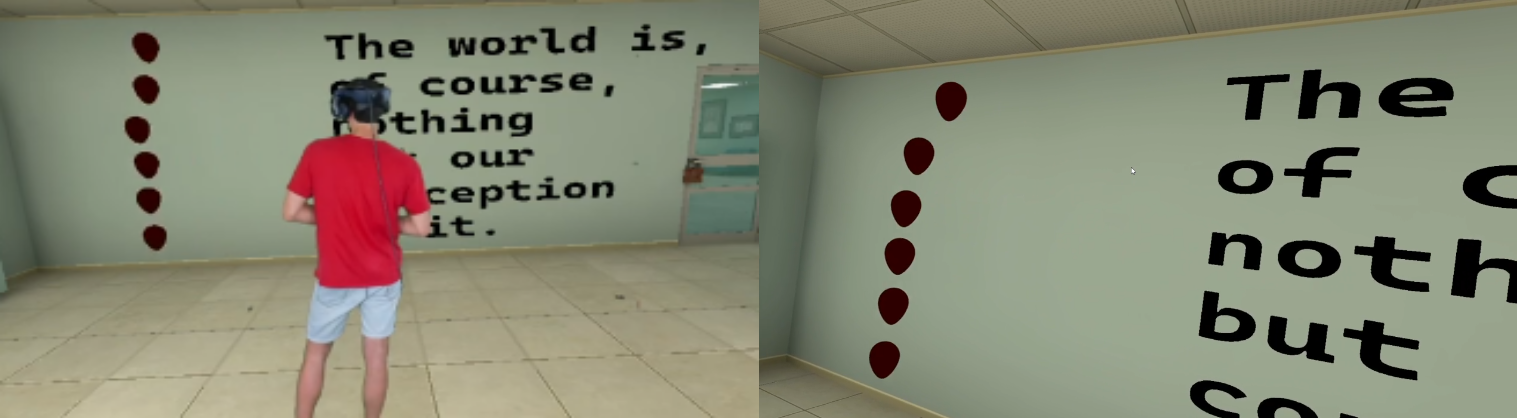}
    \caption*{\textbf{Puzzle 15}: Similarly to puzzle $14$ and inspired by screen refresh rate tests~\cite{UFO_test}, we present a number of balloons moving at different refresh rates. Depending on the refresh rate of the VR device, users cannot distinguish between some balloons.}
\end{figure}

\begin{figure}[H]
    \centering
    \includegraphics[width=\linewidth]{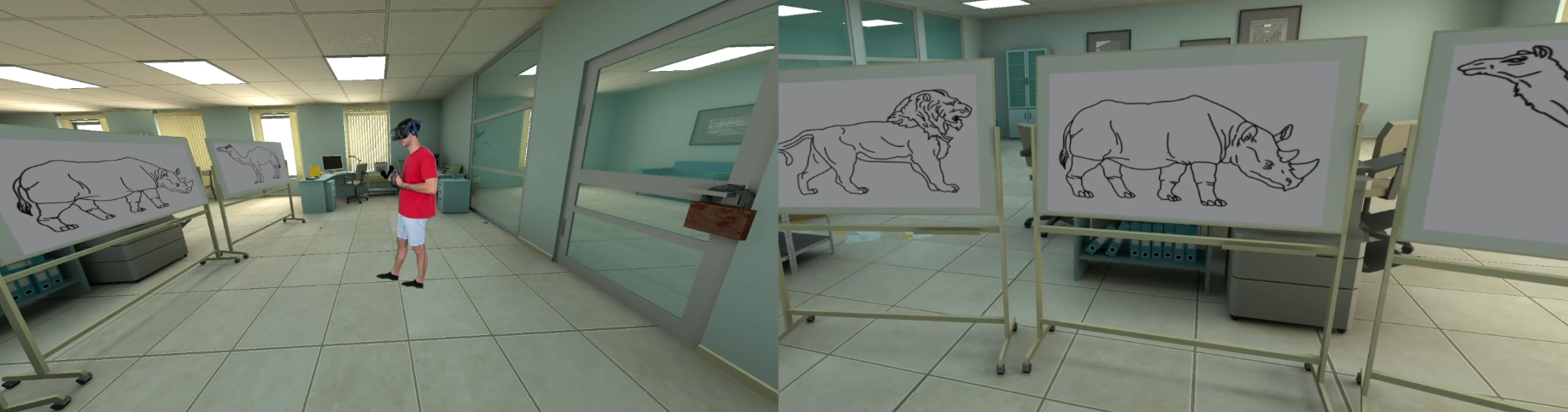}
    \caption*{\textbf{Puzzle 16}: To deploy the ``naming'' MoCA task, the room presents three whiteboards depicting three animals.}
\end{figure}

\begin{figure}[H]
    \centering
    \includegraphics[width=\linewidth]{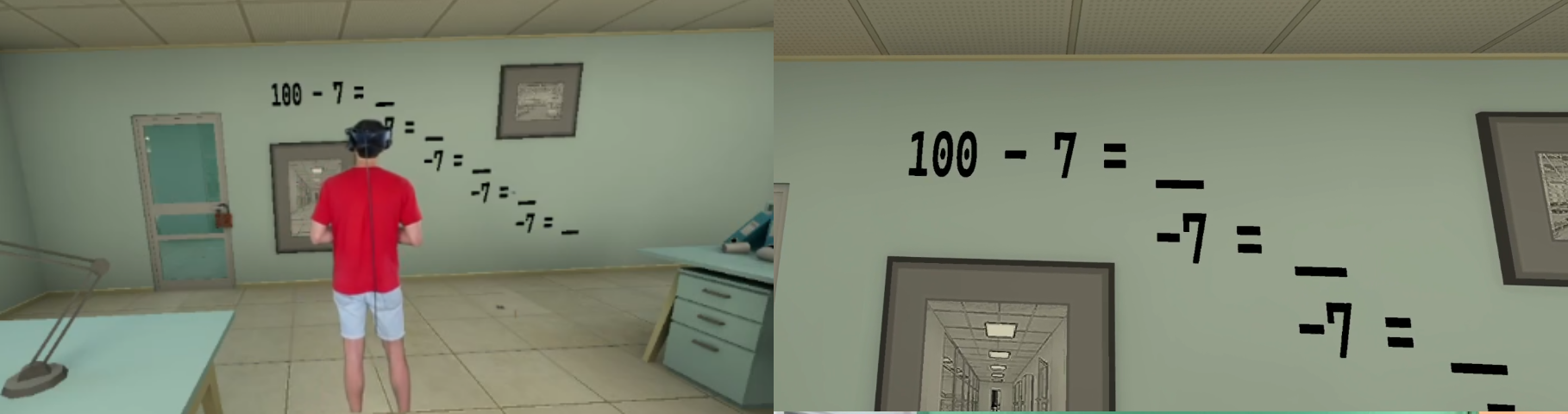}
    \caption*{\textbf{Puzzle 17}: To measure an ``attention'' task from MoCA, we present a serial seven subtraction starting at $100$, the password is the sequence of numbers that lead to the final answer: ``$65$.''}
\end{figure}

\begin{figure}[H]
    \centering
    \includegraphics[width=\linewidth]{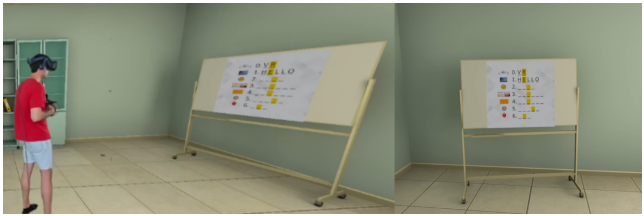}
    \caption*{\textbf{Puzzle 18}: This room contains puzzle $7$, thereby measuring delayed recall from the MoCA test.}
\end{figure}

\begin{figure}[H]
    \centering
    \includegraphics[width=\linewidth]{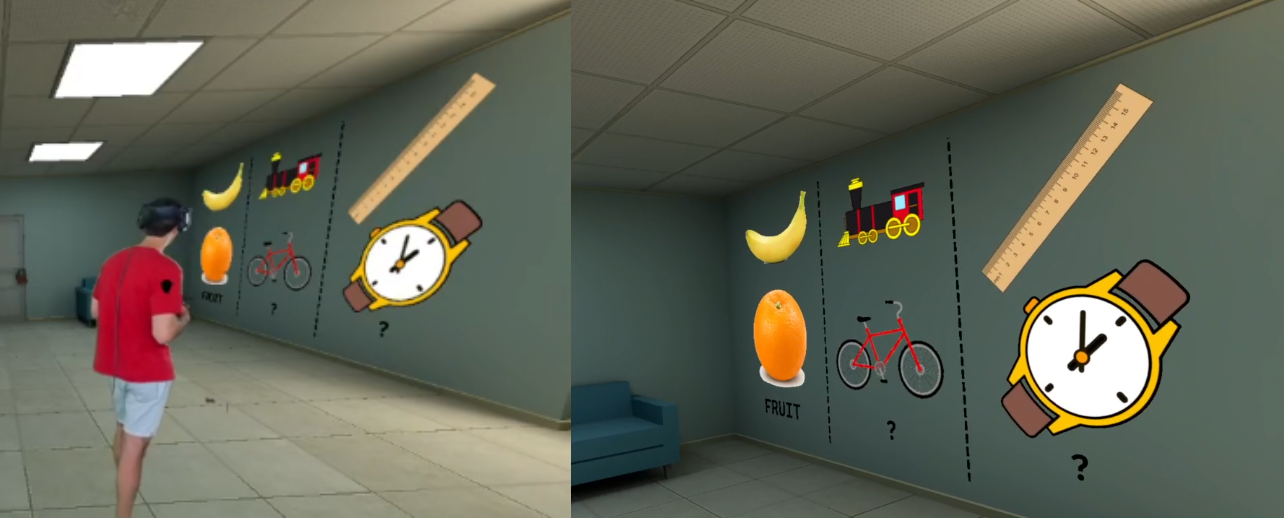}
    \caption*{\textbf{Puzzle 19}: This room pictographically recreates the MoCA abstraction test.}
\end{figure}

\noindent \textbf{Puzzle 20} (no image): To complete this audio-only room, the participant must repeat aloud two recorded sentences after listening to them once, thereby measuring one of the language tests of the MoCA.

\begin{figure}[H]
    \centering
    \includegraphics[width=\linewidth]{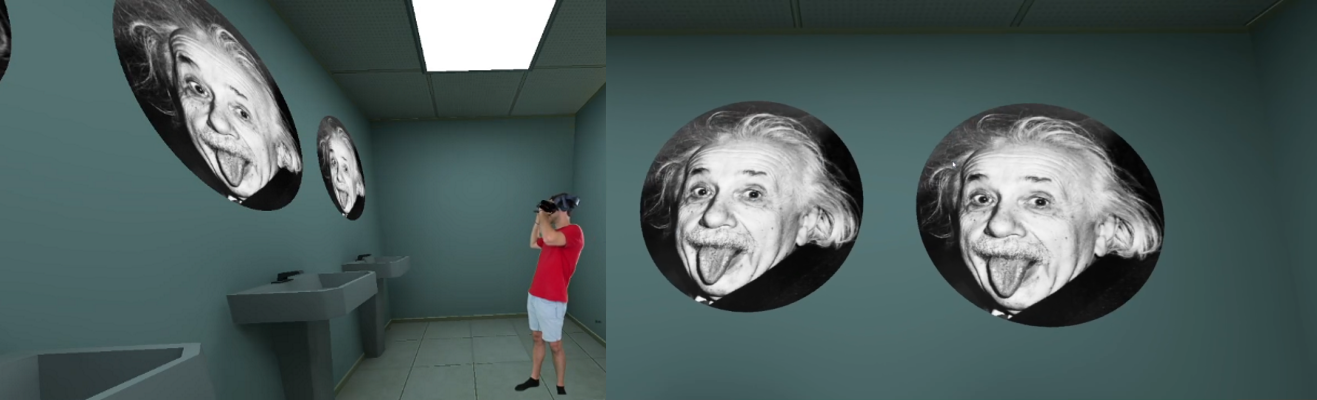}
    \caption*{\textbf{Puzzle 21}: The (noisy) room depicts three pictures of a famous physicist---``\textit{Albert Einstein}'' is the password.}
\end{figure}

\begin{figure}[H]
    \centering
    \includegraphics[width=\linewidth]{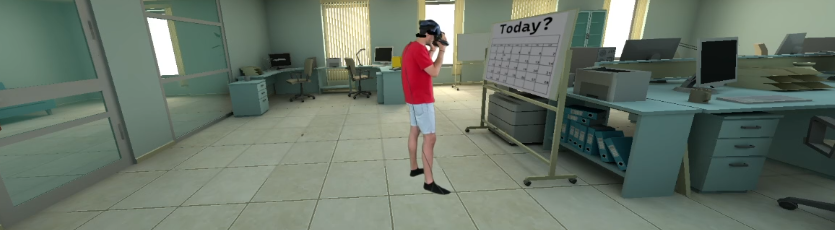}
    \caption*{\textbf{Puzzle 22}: The room presents calendar days on a whiteboard with ``\textit{Today?}'' as the header and without disclosing the year, month, weekday, or date, which prompts the participant to identify the date of the experiment, thereby measuring one variable of the orientation task in MoCA.}
\end{figure}

\begin{figure}[H]
    \centering
    \includegraphics[width=\linewidth]{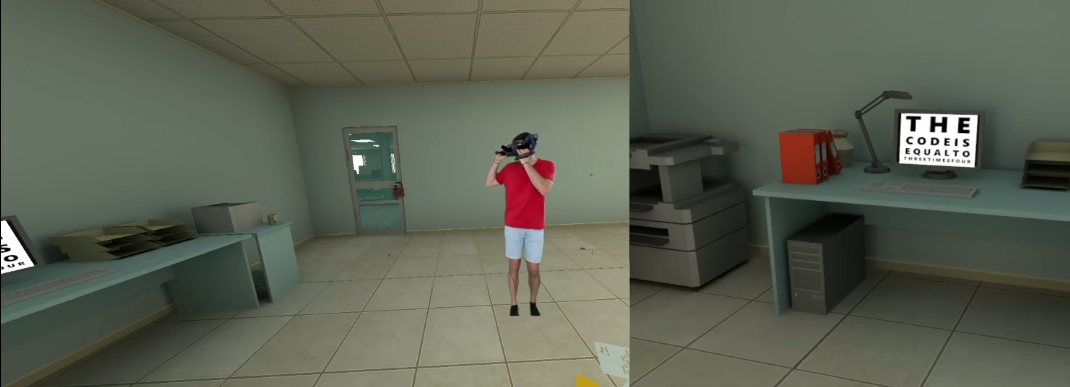}
    \caption*{\textbf{Puzzle 23}: We measure whether a participant can read the text at a close distance. We write the sentence ``\textit{The code is equal to three times four}'' in four lines on the screen of a monitor, each line becoming more diminutive than the above line.}
\end{figure}

\begin{figure}[H]
    \centering
    \includegraphics[width=\linewidth]{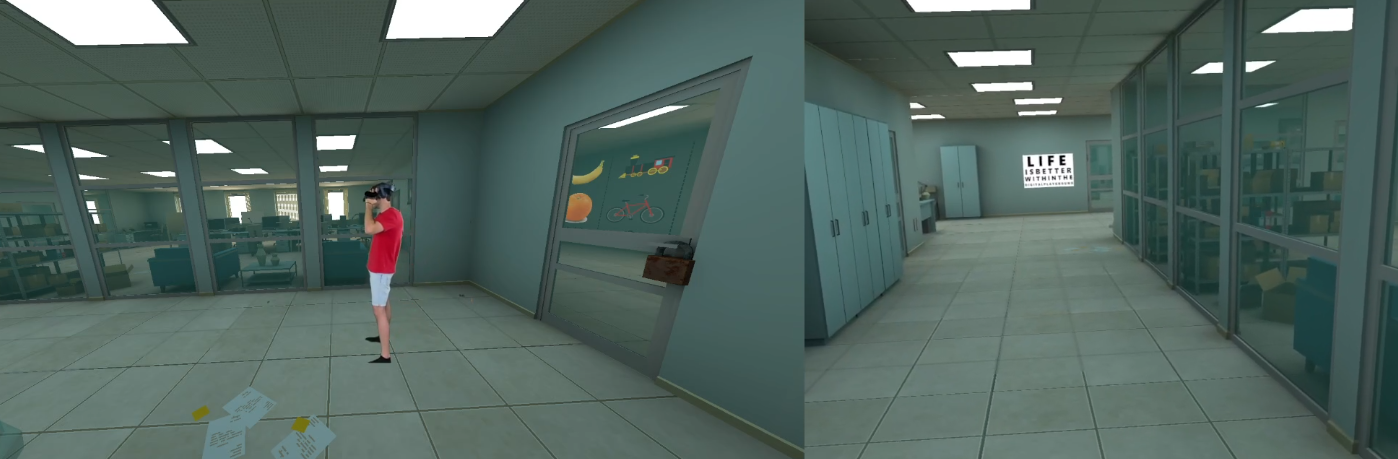}
    \caption*{\textbf{Puzzle 24}: Similarly, we measure whether a participant can read the sentence ``\textit{Life is better within the digital playground}'' at a long distance.}
\end{figure}
\clearpage
\section{Participant Distribution}
\setcounter{subsection}{0}
\label{app:distr}

\subsection{Demographics}
\noindent \textbf{Gender} \dotfill \textbf{~50} \\
Male \dotfill ~26 (52.0\%) \\
Female \dotfill ~24 (48.0\%)
\smallskip

\noindent \textbf{Age} \dotfill \textbf{~50} \\
18--23 \dotfill ~24 (48.0\%)\\
24--27 \dotfill ~20 (40.0\%)\\
28--64 \dotfill ~6 (12.0\%)
\smallskip

\noindent \textbf{Nationality} \dotfill \textbf{~50} \\
American \dotfill ~23 (46.0\%)\\
Chinese \dotfill ~8 (16.0\%)\\
Indian \dotfill ~6 (12.0\%)\\
German \dotfill ~3 (6.0\%)\\
Canadian \dotfill ~2 (4.0\%)\\
Brazilian \dotfill ~1 (2.0\%)\\
British \dotfill ~1 (2.0\%)\\
Portuguese \dotfill ~1 (2.0\%) \\
Spanish \dotfill ~1 (2.0\%) \\
Swiss \dotfill ~1 (2.0\%) \\
\textit{Undisclosed} \dotfill ~3 (6.0\%)
\smallskip

\noindent \textbf{Ethnicity} \dotfill \textbf{~50} \\
Asian \dotfill ~30 (60.0\%)\\
White \dotfill ~14 (30.0\%)\\
Black \dotfill ~3 (6.0\%)\\
Hispanic \dotfill ~3 (6.0\%)
\smallskip

\noindent \textbf{Income} \dotfill \textbf{~50} \\
$\leq$ \$25k \dotfill ~20 (40.0\%)\\
\$25k{--}\$50k \dotfill ~15 (30.0\%)\\
\$50k{--}\$100k \dotfill ~7 (14.0\%)\\
$\geq$ \$100k \dotfill ~3 (6.0\%)\\
\textit{Undisclosed} \dotfill ~5 (10.0\%)
\smallskip

\noindent \textbf{Disability Status} \dotfill \textbf{~50} \\
None \dotfill ~46 (92.0\%)\\
Mental \dotfill ~3 (6.0\%)\\
Physical \dotfill ~1 (2.0\%)
\smallskip

\noindent \textbf{Languages} \dotfill \textbf{~50} \\
Chinese \dotfill ~20 (40.0\%)\\
Spanish \dotfill ~14 (28.0\%)\\
French \dotfill ~13 (26.0\%)\\
Hindi \dotfill ~7 (14.0\%)\\
None \dotfill ~6 (12.0\%)\\
Portuguese \dotfill ~2 (4.0\%)\\
Arabic \dotfill ~1 (2.0\%)
\smallskip

\subsection{Biometrics}
\noindent \textbf{Height} \dotfill \textbf{~50} \\
150~cm -- 165~cm \dotfill ~18 (36.0\%)\\
166~cm -- 175~cm \dotfill ~16 (32.0\%)\\
176~cm -- 189~cm \dotfill ~16 (32.0\%)
\smallskip

\noindent \textbf{Wingspan} \dotfill \textbf{~50} \\
100~cm -- 169~cm \dotfill ~21 (42.0\%)\\
170~cm -- 179~cm \dotfill ~18 (36.0\%)\\
180~cm -- 191~cm \dotfill ~11 (22.0\%)
\smallskip

\noindent \textbf{Longer Arm} \dotfill \textbf{~50}  \\
Left \dotfill ~26 (52.0\%)\\
Right \dotfill ~18 (36.0\%)\\
Same \dotfill ~6 (12.0\%)
\smallskip

\noindent \textbf{Reaction Time} \dotfill \textbf{~50} \\
$>$ 250~ms \dotfill ~27 (54.0\%)\\
$<$ 250~ms \dotfill ~23 (46.0\%) 
\smallskip

\noindent \textbf{IPD} \dotfill \textbf{~50}\\
$<$ 63~mm \dotfill ~26 (52.0\%)\\
63~mm -- 66~mm \dotfill ~21 (41.0\%)\\
$>$ 66~mm  \dotfill ~3 (6.0\%)
\smallskip

\noindent \textbf{Fitness} \dotfill \textbf{~50}\\
Moderate \dotfill ~32 (64.0\%)\\
High \dotfill ~10 (20.0\%)\\
Low \dotfill ~8 (16.0\%)
\smallskip

\noindent \textbf{Colorblindness}\dotfill \textbf{~50} \\
None \dotfill ~48 (96.0\%)\\
Deuteranopia \dotfill ~2 (4.0\%)
\smallskip

\noindent \textbf{Hyperopia} \dotfill \textbf{~50}\\
None \dotfill ~28 (56.0\%)\\
Severe \dotfill ~13 (26.0\%)\\
Mild \dotfill  ~9 (18.0\%)
\smallskip

\noindent \textbf{Myopia} \dotfill \textbf{~50}\\
Severe \dotfill ~32 (66.0\%)\\
None \dotfill ~14 (28.0\%)\\
Mild \dotfill ~4 (8.0\%)
\smallskip

\noindent \textbf{MoCA}\dotfill \textbf{~50}\\
Pass ($>$ 26) \dotfill ~43 (86.0\%)\\
Fail ($\leq$ 26) \dotfill ~7 (14.0\%)
\smallskip

\noindent \textbf{Handedness}\dotfill \textbf{~50}\\
Right \dotfill ~47 (94.0\%)\\
Left \dotfill ~3 (6.0\%)
\smallskip

\subsection{Environment}
\noindent \textbf{Location} \dotfill \textbf{~50}\\
Location A \dotfill ~26 (52.0\%)\\
Location B \dotfill ~20 (40.0\%)\\
Location C \dotfill ~4 (8.0\%)
\smallskip

\noindent \textbf{Room Size} \dotfill \textbf{~50}\\
5~$m^2$--7~$m^2$ \dotfill ~24 (48.0\%)\\
$>$ 8~$m^2$ \dotfill ~20 (40.0\%)\\
$<$ 5~$m^2$ \dotfill ~4 (8.0\%)
\smallskip

\noindent \textbf{Duration} \dotfill \textbf{~50}\\
$\leq$~15~min \dotfill ~22 (44.0\%)\\
16~min--20~min \dotfill ~18 (36.0\%)\\
20~min--30~min \dotfill ~10 (20.0\%)
\smallskip

\noindent \textbf{Device} \dotfill \textbf{~50}\\
Vive Pro 2 \dotfill ~26 (52.0\%)\\ 
Oculus Quest 2 \dotfill ~21 (42.0\%)\\
HTC Vive \dotfill ~3 (6.0\%)\\
\section{Sources of Ground Truth}
\label{app:instr}

\noindent {Gender} \dotfill {Self-Reported} \\
\noindent {Age} \dotfill {Self-Reported} \\
\noindent {Nationality} \dotfill {Self-Reported} \\
\noindent {Ethnicity} \dotfill {Self-Reported} \\
\noindent {Income} \dotfill {Self-Reported} \\
\noindent {Disability Status} \dotfill {Self-Reported} \\
\noindent {Languages} \dotfill {Self-Reported}\\
\noindent {Height} \dotfill {Stadiometer} \\
\noindent {Wingspan} \dotfill {Measuring Tape} \\
\noindent {Longer Arm} \dotfill {Measuring Tape}  \\
\noindent {Reaction Time} \dotfill {Application} \\
\noindent {IPD} \dotfill {Pupilometer}\\
\noindent {Fitness} \dotfill {Self-Reported}\\
\noindent {Colorblindness}\dotfill {Self-Reported} \\
\noindent {Hyperopia} \dotfill {Self-Reported}\\
\noindent {Myopia} \dotfill {Self-Reported}\\
\noindent {MoCA}\dotfill {Administered}\\
\noindent {Handedness}\dotfill {Self-Reported}\\
\noindent {Location} \dotfill {GPS}\\
\noindent {Room Size} \dotfill {Measuring Tape}\\
\noindent {Duration} \dotfill {Chronometer}\\
\noindent {Device} \dotfill {Observed}\\

\clearpage

\end{document}